\documentclass[10pt,journal,compsoc]{IEEEtran}
\ifCLASSOPTIONcompsoc
  \usepackage[compress]{cite}
\else
  \usepackage{cite}
\fi

\usepackage{ragged2e}
\usepackage{underscore}
\usepackage{adjustbox}
\usepackage{hyperref}
\usepackage{amssymb}
\usepackage{graphicx}
\usepackage{multirow}
\usepackage{bbding}
\usepackage{float}
\usepackage{cancel}
\usepackage{url}
\usepackage{graphicx}
\usepackage{color}
\usepackage[linesnumbered, ruled]{algorithm2e}
\usepackage{algorithm2e,setspace}
\usepackage{stfloats}
\usepackage{multirow}
\usepackage{makecell}
\usepackage{multicol}
\usepackage[noend]{algpseudocode}
\usepackage{bm}
\usepackage{float}
\usepackage[table,xcdraw]{xcolor}
\usepackage[normalem]{ulem}

\useunder{\uline}{\ul}{}
\usepackage{float}  
\usepackage{lipsum}


\ifCLASSINFOpdf
\else
\fi
\begin{document}
%\title{Knowledge Distillation Meets Federated Multi-task Learning in Mobile Edge Computing}

%\title{FedCache: A Novel Federated Learning Architecture for Personalized Edge Intelligence}
%\title{Personalized Edge Intelligence with Cache-based Knowledge Retrieval}
%\title{FedCache: A Novel Federated Learning Architecture for Personalized Edge Intelligence}
%\title{Federated Learning with Server-Side Knowledge Caching and Retrieval: An Architecture for Personalized Edge Intelligence}
%\title{FedCache: A Novel Personalized Federated Learning Architecture with Knowledge Caching and Retrieval in the Edge}
\title{FedCache: A Knowledge Cache-driven \\Federated Learning Architecture for Personalized Edge Intelligence}

\author{Zhiyuan~Wu,~\IEEEmembership{Member,~IEEE,}
        Sheng~Sun,
        Yuwei~Wang,~\IEEEmembership{Member,~IEEE,}
        Min~Liu,~\IEEEmembership{Senior~Member,~IEEE,}\\
        Ke~Xu,~\IEEEmembership{Senior~Member,~IEEE,}
        Wen~Wang,
        Xuefeng~Jiang,~Bo~Gao,~\IEEEmembership{Member,~IEEE},
        %Tianliu~He,
        and~Jinda Lu
\IEEEcompsocitemizethanks{\IEEEcompsocthanksitem Zhiyuan Wu, Wen Wang, and Xuefeng Jiang are with the Institute of Computing Technology, Chinese Academy of Sciences, Beijing, China, and also with the University of Chinese Academy of Sciences, Beijing, China.
E-mails: \{wuzhiyuan22s, wangwen22s, jiangxuefeng21b\}@ict.ac.cn. %tianliu.he@foxmail.com.
\IEEEcompsocthanksitem Sheng Sun and Yuwei Wang are with the Institute of Computing Technology, Chinese Academy of Sciences, Beijing, China.
E-mails: \{sunsheng, ywwang\}@ict.ac.cn.
\IEEEcompsocthanksitem Min Liu is with the Institute of Computing Technology, Chinese Academy of Sciences, Beijing, China, and also with the Zhongguancun Laboratory, Beijing, China.
E-mail: liumin@ict.ac.cn
\IEEEcompsocthanksitem Ke Xu is with the Department of Computer Science and Technology, Tsinghua University, Beijing, China, and also with the Zhongguancun Laboratory, Beijing, China.
E-mail: xuke@tsinghua.edu.cn.
\IEEEcompsocthanksitem Bo Gao is with the School of Computer and Information Technology, and the Engineering Research Center of Network Management Technology for High-Speed Railway of Ministry of Education, Beijing Jiaotong University, Beijing, China.
E-mail: bogao@bjtu.edu.cn.
\IEEEcompsocthanksitem Jinda Lu is with the School of Information Science and Technology, University of Science and Technology of China, Hefei, China.
E-mail: lujd@mail.ustc.edu.cn.
\IEEEcompsocthanksitem Corresponding author: Yuwei Wang.
}% <-this % stops an unwanted space
\thanks{
	%Manuscript received January xxx, xxx; revised xxx, xxx, xxx and xxx, xxx, xxx; accepted xxx, xxx, xxx. Date of publication xxx, xxx, xxx; date of current version ???, ???, ???. 
	This work was supported by the National Key Research and Development Program of China (No. 2021YFB2900102), the National Natural Science Foundation of China (No. 62072436), the Innovation Capability Support Program of Shaanxi (No. 2023-CX-TD-08), Shaanxi Qinchuangyuan "scientists+engineers" team (No.2023KXJ-040), and the Innovation Funding of ICT, CAS (No.E261080). }
}

\markboth{accepted by IEEE Transactions on Mobile Computing}%
% \markboth{{\tiny This work has been submitted to the IEEE for possible publication. Copyright may be transferred without notice, after which this version may no longer be accessible. }}
{Shell \MakeLowercase{\textit{et al.}}: Bare Demo of IEEEtran.cls for Computer Society Journals}




\IEEEtitleabstractindextext{%
\begin{abstract}
\justifying
%Federated Learning (FL) enables privacy-preserving distributed model training over devices.
%However, most existing FL methods are based on FedAvg, which suffer from unaffordable communication burden caused by large-scale parameters transmission between clients and the server.
%To address this issue, we propose a communication-efficient and accuracy-guaranteed FL architecture, named FedRC. With the retrieval-based knowledge distillation technique during training,  clients can match the most relevant ensembled knowledge in the server-side knowledge cache and use it for distillation on local models, so as to learn from other clients in a communication-efficient manner. 
%Edge Intelligence (EI) provides real-time data analysis and decision-making close to data sources, raising privacy concerns due to uploading raw data on devices. 
%Edge intelligence (EI) pushes artificial intelligence (AI) to the edge for providing real-time data analysis and decision-making close to data sources.
Edge Intelligence (EI) allows Artificial Intelligence (AI) applications to run at the edge, where data analysis and decision-making can be performed in real-time and close to data sources.
To protect data privacy and unify data silos distributed among end devices in EI, Federated Learning (FL) is proposed for collaborative training of shared AI models across multiple devices without compromising data privacy.
However, the prevailing FL approaches cannot guarantee model generalization and adaptation on heterogeneous clients.
%However, FL does not guarantee that the trained model is generalizable and adaptable to all clients.
%while keeping on-devices data decentralized.
%Federated Learning (FL) is a privacy-preserving distributed learning paradigm in EI, but the learned uniform model in conventional FL is difficult to deploy and generalize to all devices. 
%Personalized Federated Learning (PFL) enables differentiated model training to achieve personalized services in EI,whereas prevalent PFL architectures suffer from the homogeneity of on-device model architectures as well as the communication burden due to large-scale parameters transmission and aggregation. 
Recently, Personalized Federated Learning (PFL) has drawn growing awareness in EI, as it enables a productive balance between local-specific training requirements inherent in devices and global-generalized optimization objectives for satisfactory performance.
However, most existing PFL methods are based on the Parameters Interaction-based Architecture (PIA) represented by FedAvg, which suffers from unaffordable communication burdens due to large-scale parameters transmission between devices and the edge server.
In contrast, Logits Interaction-based Architecture (LIA) allows to update model parameters with logits transfer and gains the advantages of communication lightweight and heterogeneous on-device model allowance compared to PIA. Nevertheless, previous LIA methods attempt to achieve satisfactory performance either relying on unrealistic public datasets or increasing communication overhead for additional information transmission other than logits.
%or increasing communication overhead due to the transmission of non-negligible information other than logits.
To tackle this dilemma, we propose a knowledge cache-driven PFL architecture, named FedCache,
%which is the first sample-grained logits interaction-based PFL architecture without public datasets.
%to enable sample-grained logits interaction between clients and the server without relying on public datasets.
%Specifically, FedCache 
which reserves a knowledge cache on the server for fetching personalized knowledge from the samples with similar hashes to each given on-device sample.
During the training phase, ensemble distillation is applied to on-device models for constructive optimization with personalized knowledge transferred from the server-side knowledge cache.
Empirical experiments on four datasets demonstrate that FedCache achieves comparable performance with state-of-art PFL approaches, with more than two orders of magnitude improvements in communication efficiency. 
Our code and DEMO are available at \textit{\url{https://github.com/wuzhiyuan2000/FedCache}}.
\end{abstract}
\begin{IEEEkeywords}
Distributed architecture, edge computing, personalized federated learning, knowledge distillation, communication efficiency
% , information retrieval
\end{IEEEkeywords}
}
\maketitle

% - because all conference papers position the abstrac like regular
\IEEEdisplaynontitleabstractindextext

% \fi

\IEEEpeerreviewmaketitle
\vspace{20pt}
\IEEEraisesectionheading{\section{Introduction}}
%\IEEEPARstart{F}{ederated} Learning (FL) is an emerging machine learning paradigm for training models on private data distributed over devices. Traditional FL bases on the architecture of FedAvg \cite{mcmahan2017communication,reddi2021adaptive,li2020federated}, and the communication overhead from periodically large-scale parameter transmission is non-negligible for devices \cite{wu2022communication,sattler2021cfd}. 
%Although a series of approaches enables training without parameters transmission of models on clients, they either impractically rely on an public dataset \cite{itahara2021distill}, require transmissing embedded features or generators of non-negligible size  \cite{wu2023fedict,he2020group,zhu2021dataicml}, or achieves inferior performance than early FL algorithms represented by FedAvg \cite{jeong2018communication,sattler2021cfd}.
%\IEEEPARstart{E}{dge} intelligence is an emerging distributed computing technology that integrates edge computing and artificial intelligence to achieve real-time data analysis and intelligent decision-making on devices by deploying machine learning algorithms to devices close to data sources \cite{xu2021edge}. 
\IEEEPARstart{E}{dge} Intelligence (EI) is an emerging technology for the marriage of edge computing and Artificial Intelligence (AI), enabling real-time data analysis and decision-making close to data sources instead of relying solely on the cloud \cite{xu2021edge}.
With the proliferation of mobile devices and the unprecedented amount of data generated by ubiquitous devices, EI is playing an increasingly important role in many areas such as unmanned vehicles \cite{yang2022edge}, smart homes \cite{nasir2022enabling}, recommender systems \cite{gong2020edgerec}, etc. 
However, conventional centralized EI paradigms require uploading raw data for training pervasive AI models, raising privacy concerns about sensitive data leakage.

\begin{figure}[t]
	\centering
	\includegraphics[width=0.5\textwidth]{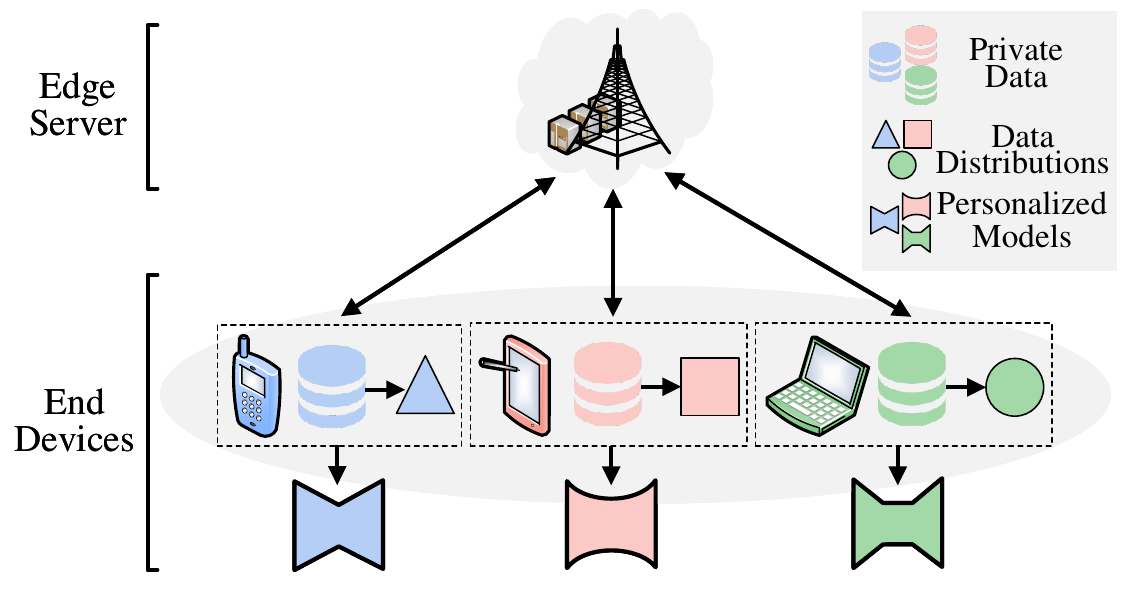}
	\caption{\textcolor{black}{Schematic diagram of personalized federated learning for edge intelligence.}}
	\label{ei}
\end{figure}

Federated Learning (FL) is a privacy-preserving distributed learning paradigm that enables multiple data owners to collaboratively train AI models without sharing owners' private data.
Due to the benefits of data localization and privacy protection, FL has shown great potential in various EI applications, such as healthcare \cite{antunes2022federated}, smart transportation \cite{liu2023online}, industrial manufacturing \cite{kanagavelu2021federated}, etc.
Unfortunately, the prevailing FL approaches \cite{mcmahan2017communication,li2020federated} require all participating devices (named clients) to share a uniform model, which is extremely difficult to deploy and generalize to all devices because of the inherent characteristics of device variation in terms of data heterogeneity, resources limitation, task differentiation, etc \cite{kulkarni2020survey,tan2022towards}.
Recent studies pay much attention to Personalized Federated Learning (PFL) \cite{tan2022towards} for addressing the differential training challenges in EI via building personalized models for individual devices, \textcolor{black}{as shown in Fig. \ref{ei}}.
However, most PFL approaches \cite{t2020personalized,mills2021multi,jin2022personalized} adopt the Parameters Interaction-based Architecture (PIA) represented by FedAvg \cite{mcmahan2017communication}, which requires homogeneity of on-device model architectures and imposes tremendous communication burden caused by large-scale parameters exchange between clients and the server for bandwidth-limited devices \cite{wu2022communication,sattler2021cfd}.

Furthermore, by applying knowledge distillation technology \cite{hinton2015distilling,wu2021spirit,gou2021knowledge} to PFL, a series of communication-lightweight and heterogeneous model-allowable PFL architectures with logits (usually called knowledge) exchange instead of interacting model parameters are put forward.
These architectures, which we call Logits Interaction-based Architecture (LIA), bring the benefits of saving orders of magnitude of communication overhead and training models with heterogeneous architectures.
Related literatures \cite{zhang2021parameterized,jee2023communication,wu2023fedict,jeong2018communication,sattler2021cfd} fall into two types of architectures based on the granularity of the interacted logits during training: Class-grained Logits Interaction-based Architecture (CLIA) and Sample-grained Logits Interaction-based Architecture (SLIA). 
Thereinto, SLIA is drawn more attention since it allows for fine-grained interaction of logits for performance guarantee.
However, existing methods based on SLIA endeavor to achieve satisfactory performance either relying on additional client-side training on unrealistic public datasets \cite{zhang2021parameterized,jee2023communication}, or requiring the transfer of embedded features with non-negligible sizes in addition to logits \cite{wu2023fedict,wu2022exploring}.
They appear to be unfriendly to devices by reason of the induction of intensive computation, tremendous communication, or public datasets reliance, making them unsuitable for practical applications in EI \cite{wu2023survey}.
%most researches endeavor to achieve satisfactory training performance based on sample-grained logits interaction either relying on unrealistic public dataset  \cite{zhang2021parameterized,jee2023communication} or requiring the transfer of embedded features or generators with non-negligible sizes \cite{wu2023fedict,zhu2021dataicml}, undoubtedly adverse to practical application at the edge. A little of literatures [] attempt to eliminate the above-mentioned supplementary limitations via class-grained logits interaction, paying the price of accuracy degradation.
%Although a number of approaches \cite{zhang2021parameterized,jee2023communication,wu2023fedict,zhu2021dataicml,jeong2018communication,sattler2021cfd} avoid transmitting model parameters during training, they compromise the performance or feasibility of deploying in EI. Specifically, they either have inferior performance than early FL algorithms \cite{jeong2018communication,sattler2021cfd}, rely on an unrealistic public dataset \cite{zhang2021parameterized,jee2023communication}, or require the transfer of embedded features or generators of non-negligible sizes \cite{wu2023fedict,zhu2021dataicml}. As none of the above approaches can practically address the challenge of personalization in EI, a novel FL architecture is urgently needed.
%to balance the practical needs in EI scenarios.

In this paper, we develop a novel device-friendly PFL architecture that is suitable for EI, named knowledge cache-driven FL architecture (FedCache).
\textcolor{black}{FedCache is a novel client-server interaction paradigm,} which maintains a knowledge cache on the server to store the latest knowledge associated with each private sample, and applies a customized knowledge cache-driven personalized distillation technique for on-device model training. During the initialization process, all private samples on clients are encoded into hashes via a deep pre-trained neural network, so as to discern the relational degree among samples in a privacy-preserving manner.
During the training process, each on-device model is optimized via personalized knowledge distillation over the ensemble of relevant knowledge whose corresponding hashes are the $R$-nearest neighbors of the hash of the given sample to be optimized on, which is fetched from the server-side knowledge cache. 
To our best knowledge, FedCache is \textbf{the first Sample-grained Logits Interaction-based Architecture (SLIA) dispensed with features transmission and public datasets}, ensuring the satisfactory performance of on-device models while meeting the practical limitations of EI.
% where related knowledge is fetched from the server-side knowledge cache whose hashes are $R$-nearest neighbors of the hash of the sample to be optimized on.
%With the aforementioned training procedure, clients can efficiently learn from each other with personalized ensemble knowledge transfer.

%Each sample of private data retrieves grouped personalized knowledge (logits) in the server-side knowledge cache, where the samples from which the knowledge is extracted have a similar hash value to the private sample.
%We leverage a deep, pre-trained neural network to encode all private samples on clients, retrieving grouped personalized knowledge with HNSW \cite{malkov2018efficient} that matches the best as measured by a customized hash similarity function.
%Based on the fetched knowledge from the knowledge cache, we apply ensemble distillation to corresponding on-device models, so that 
%clients can efficiently learn from each other with personalized knowledge transfer.

%In general, our proposed FedCache architecture integrates the following superiorities in EI.
In general, we summarize the contributions of our proposed FedCache as follows:
\begin{itemize}
\item
\textbf{Device Friendliness.} FedCache is a device-friendly architecture that enables only small-scale ensemble logits to be transferred between clients and the server during training without needing public datasets. Meanwhile, FedCache supports collaborative training on devices with heterogeneous models.
\item 
\textbf{Scalability.} FedCache is a highly scalable architecture for large-scale devices since it eliminates the need to keep a cumbersome global model on the server and also enables asynchronous training, effectively reducing the server-side computation and client-server synchronization consumption.
\item 
\textbf{Effectiveness.} FedCache is compared with state-of-art PFL methods with various architectures on four common datasets. 
Results confirm that FedCache achieves performance comparable to benchmark algorithms while improving communication efficiency by two orders of magnitude.
\end{itemize}

%To verify the effectiveness of FedCache, we compared it with state-of-art PFL methods with various architectures on four common datasets. Results confirm that FedCache achieves comparable performance with benchmark algorithms but with two orders of magnitude improvement in communication efficiency.
%\begin{figure}[!t]
%	\begin{minipage}[]{0.45\linewidth}
%		\centering
%		\includegraphics[width=6cm]{framework.pdf}
%		\caption{Architecture of FedRC. \\(1) Upload coded sample. \\(2) Build relations. \\(3) Upload knowledge. \\(4) Retrieve and ensemble knowledge. \\(5) Download ensembled knowledge. \\(6) Distillation.}
%		\label{arch}
%	\end{minipage}%
%	\begin{minipage}[]{0.45\linewidth}
%		\centering
%		\includegraphics[width=5.5cm]{sim.pdf}
%		\caption{Matched samples with $R=3$. With hash-based similarity retrieval, very similar samples can be matched, based on which knowledge from other models can be transferred. XXXXXX}
%		\label{sim}
%	\end{minipage}
%\end{figure}

\section{Related Work}

\subsection{Federated Learning for Personalized Edge Intelligence}
As a uniform shared model cannot accommodate multiple clients with diverse tasks and capabilities \cite{tan2020federated,kulkarni2020survey}, personalization techniques in FL are needed to adapt clients-side individualized requirements.
Specifically, federated multi-task learning in edge computing \cite{wu2023fedict,mills2021multi} allows clients to train personalized neural networks to accommodate the differentiated data distribution in their respective tasks. \cite{jin2022personalized} retains historical personalized models on devices, allowing current models to distill knowledge from previous models for personalized EI. 
\cite{zhou2022sourcetarget} leverages a source-free unsupervised domain adaptation approach to adapt large source-domain models to target data on devices, while adopting lite residual hypothesis transfer to save the storage overhead during the adaptation process.
In addition, \cite{jiang2020customized} considers the personalization of accuracy targets on clients and uses adaptive learning rates to allow clients that reach the target to exit in advance for saving resources.

Unlike prior works, we focus on the architecture design of PFL to enable efficient communication as well as asynchronous training, while also supporting heterogeneous on-device models without requiring public datasets, aiming to bridge the gap between PFL and practical applications in EI.

\subsection{Knowledge Distillation in Federated Edge Learning}
As noted in \cite{wu2023survey}, knowledge distillation has emerged as an important technique for addressing challenges in federated edge learning.
\cite{itahara2021distill,jeong2018communication} leverage knowledge distillation as a communication protocol for exchanging model representations among devices and the edge server, enabling communication-efficient training over heterogeneous models. 
\cite{zhang2022fedzkt,yu2022resource} transfer knowledge from edge models to differentiated on-device models in FL under device heterogeneity.
\cite{wu2023fedict,jin2022personalized} conduct distillation-based personalized FL optimization over on-device models to support personalized EI.

%However, the above approaches cannot simultaneously balance the need of differentiating devices, personalized training, no public data requirement, asynchronous training, and high communication efficiency. In contrast, our proposed FedCache architecture fulfills all of these requirements for personalized EI.

However, the above approaches cannot achieve satisfactory performance in a device-friendly manner, since they either need features/model parameters transfer or rely on public data during the training process.
%simultaneously balance the need of differentiating devices, personalized training, no public data requirement, asynchronous training, and high communication efficiency. 
In contrast, our FedCache architecture solves all the above-mentioned problems and enables practical use in personalized EI.

\section{Preliminary and Motivation}
\subsection{Background and Notations}
We investigate PFL for EI, where distributed devices (named clients) collaboratively train $C$-class classification models coordinated by an edge server (named the server) while keeping private data on devices.
We assume that $K$ clients participate in PFL, and each client $k \in \{1,2,...,K\}$ occupies a private dataset ${{\cal D}}^k := \bigcup\limits_{i = 1}^{{N^k}} {\{ ({{X}_i^k},{{y}_i^k})\} }$, where $N^k$ is the number of samples in ${\cal D}^k$, and $X^k_ i$, $y^k_i$ are the $i$-th data and label in ${\cal D}^k$, respectively. 
%\textcolor{black}{Moreover, $L_{CE}(\cdot)$ denotes the cross-entropy loss, and $KL(\cdot)$ denotes the Kullback-Leibler divergence loss.}
Each device $k$ owns a personalized model $M^k:=(W^k,f^k)$ \textcolor{black}{with possible different model parameters or architectures,} where $W^k$ is the model parameters of $M^k$ and $f^k(\cdot)$ is the non-linear mapping determined by $M^k$.
The goal of each device is to improve the User model Accuracy (UA) \cite{mills2021multi} of its personalized model on its private data as much as possible.
The optimization objective of the PFL system is to maximize the Maximum Average UA (MAUA) of all clients, that is to achieve generally satisfactory performance on each client. 
\textcolor{black}{The main notations and descriptions can be referred to TABLE \ref{notation}.}

\begin{table}[!t]
	\centering
	\caption{Main notations with descriptions.}
\renewcommand\arraystretch{1.2}
	\begin{tabular}{c|c}
		\hline
		\textbf{Notation} & \textbf{Description} \\ \hline
		$C$        & The number of classes               \\
        $K$        & The number of clients               \\
        $\mathcal{D}^k$      & The local dataset of client $k$     \\
		${X}_i^k$      & The $i$-th sample of $\mathcal{D}^k$     \\
		${y}_i^k$      & The label of ${X}_i^k$     \\
          $N^k$      & The number of samples in $\mathcal{D}^k$     \\
          $M^k$     & The personalized model on client $k$ \\
          $W^k$ & The model parameters of $M^k$ \\
          $f^k$     & The non-linear mapping determined by $M^k$ \\
          $f^h$     & \begin{tabular}[c]{@{}c@{}}The hash mapping determined by a \\pre-trained deep neural network\end{tabular} \\
          $L_{CE}$     & The cross-entropy loss \\
          $KL$     & The Kullback-Leibler divergence loss \\
        $\sigma_0$     & The softmax mapping \\
          $LI$     & The label-to-index pairs \\
          $IK$     & The index-to-knowledge pairs \\
          $IH$     & The index-to-hash pairs \\
          $IR$     & The index relations pairs \\
          $KC$     & The knowledge cache \\
          $R$     & The number of related samples in $KC$\\
          $h_i^k$     & The hash value of $X_i^k$ \\
          $z_i^k$     & The knowledge of $X_i^k$ \\
          $(zr^k_i)_s$     & \begin{tabular}[c]{@{}c@{}}The $s$-th knowledge fetched for the given \\sample index $(k, i)$\end{tabular} \\
        $\overline {zr} _i^k$     & \begin{tabular}[c]{@{}c@{}}The ensembled  fetched knowledge for the \\ given sample index $(k, i)$\end{tabular} \\
        $J^k$     & The optimization objective of $M^k$ \\     
		\hline
	\end{tabular}
	\label{notation}
\end{table}

\color{black}
\subsection{Practical Limitations in Edge Intelligence}
We summarize the main practical limitations that PFL architectures need to overcome when deploying in EI:
\begin{itemize}
    \item 
    \textbf{Device Heterogeneity.} Considering the different hardware configurations of end devices such as central processing units, memory resources, and energy status, personalized models need to be adopted among devices to fit their specific characteristics \cite{tak2020federated,yu2021toward}.
    \item 
    \textbf{Communication Efficiency.} Due to the limited wireless network bandwidth between edge servers and end devices, they are not capable of large-scale communication \cite{tak2020federated,sattler2021cfd}.
    \item 
    \textbf{Data Privacy.} Devices are reluctant to share their local data with edge servers because of privacy concerns or data protection regulations \cite{yang2019federated,hoofnagle2019european}. Hence, it is difficult to obtain information about users' local data.
    \item
    \textbf{Asynchronous Optimization.} The high synchronization overhead caused by varying computation tasks, capabilities, and communication delays of different devices impedes model update \cite{nguyen2022federated,xie2019asynchronous}.
    %The update of models is hindered by high synchronization overhead as a result of varying computation tasks, capabilities, and communication delays of different devices \cite{nguyen2022federated,xie2019asynchronous}.
    \end{itemize}
\color{black}
%Considering task differentiation and hardware-related constraints such as CPU, storage as well as communication capabilities, 
%Model Hetero
%Comm Efficiency
%Public Data
%Asynchron Optimzation

\subsection{Overview of PFL Architectures}
\subsubsection{PFL Architecture based on Parameters Interaction}
For Parameters Interaction-based Architecture (PIA), each client periodically uploads locally-trained model parameters to the server and updates the local model with the server-downloaded model parameters obtained from aggregating local models.
In PFL with PIA, clients tend to upload only part of its model parameters to preserve local adaptation capabilities \cite{mills2021multi,jin2022personalized}.
\textcolor{black}{Therefore, filtered parameters aggregation weighted by local sample numbers is performed on the server side, that is:}
\begin{equation}
    {W^*} = \frac{{{N^k}}}{{\sum\limits_{l = 1}^K {{N^l}} }} \cdot filter({W^k}),
\end{equation}
where $filter(\cdot)$ filters out partial on-device model parameters to be uploaded to the server, and $W^*$ represents the aggregated model parameters on the server.

\textcolor{black}{Although PIA can preserve the personalization capabilities of on-device models by filtering model parameters, transmitting large-scale parameters for aggregation is still too costly for devices with limited communication resources} \cite{tak2020federated,sattler2021cfd}. 
\textcolor{black}{Moreover, PIA demands a high degree of homogeneity among on-device model architectures during the aggregation process, which is hard to achieve in edge intelligence scenarios where heterogeneous devices with various hardware-related constraints are prevalent \cite{diao2021heterofl,lim2020federated}.}
%Besides, PIA requires strong homogeneity of on-device model architectures during the aggregation process, which is difficult to deploy in edge intelligence scenarios where heterogeneous devices with differentiated hardware-related constraints are ubiquitous \cite{diao2021heterofl,lim2020federated}.

\subsubsection{PFL Architecture based on Logits Interaction}
For Logits Interaction-based Architecture (LIA), each client performs distillation-based optimization on the global logits downloaded from the server, without parameters transmission during training \cite{jeong2018communication,he2020group,wu2023fedict,wu2022exploring,li2019fedmd,itahara2021distill}. 
Depending on the granularity of interacted logits, existing PFL architectures can be divided into two categories: class-grained logits interaction and sample-grained logits interaction.

\noindent
\textbf{\textit{1) Class-grained Logits Interaction-based Architecture (CLIA).}}
For CLIA, the output of each sample $X^k_i$ from client $k$ needs to approach the global average logits calculated by all samples with the same label $y^k_i$ from all other clients except client $k$ \cite{jeong2018communication}, that is:
%For CLIA, each output of $X^k_i$ from client $k$ needs to partially approach uniform global per-label average logits whose corresponding label is the same as $y^k_i$, that is:
\begin{equation}
	\begin{array}{l}
	\mathop {\arg \min }\limits_{{W^k}} \sum\limits_{(X_i^k,y_i^k) \in {{\cal D}^k}} {[{L_{CE}}({\sigma _0}({f^k}(X_i^k)),y_i^k)}  \\
    \; \; \; \; \; \; \; \; \; \; \; \;+\gamma  \cdot {L_{CE}}({\sigma _0}({f^k}(X_i^k)),{\sigma _0}(\frac{{\sum\limits_{l = 1}^K {{F^{l,y_i^k}}}  - {F^{k,y_i^k}}}}{{K - 1}}))],
	\end{array}
\end{equation}
where $\sigma_0(\cdot)$ is the softmax mapping, $\gamma$ is the distillation weighting factor, and $L_{CE}(\cdot)$ denotes the cross-entropy loss. $F^{l,y_i^l}$ is the average logits calculated by the samples with the same label $y_i^l$ in client $l$, i.e.
\begin{equation}
	{F^{l,y_i^l}} = \mathop E\limits_{(X_i^k,y_i^l) \in {{\cal D}^k} \wedge y_i^l = y_i^k} {f^k}(X_i^k).
\end{equation}

Although CLIA supports model heterogeneity with lightweight communication, it only enables $C$ types of logits to be learned by each client. As clients learn very little additional server-side information compared to standalone, this PFL design is prone to reaching a performance limit.

\noindent
\textbf{\textit{2) Sample-grained Logits Interaction-based Architecture (SLIA).}}
For SLIA, the number of logits learned by on-device models are related to the number of samples \cite{he2020group,wu2022exploring,wu2023fedict,itahara2021distill,li2019fedmd}. 
Such architecture generally requires inevitable compromises of importing public datasets or increasing communication overhead, and can be classified into two forms.
\begin{itemize}
    \item 
    \textbf{SLIA with Features Exchange (SLIA-FE).} In SLIA-FE, the model parameters of client $k$ are divided into the feature extractor part $W_e^k$ and the predictor part $W_p^k$, where the prediction mapping of the feature extractor is denoted as $f_e^k(\cdot)$.
The server keeps only a large-scale classifier $W^S$ with the corresponding prediction mapping $f^S(\cdot)$.
Typically, the model on the server is updated with a linear combination of cross-entropy loss $L_{CE}(\cdot)$ and Kullback-Leibler divergence loss $KL(\cdot)$ depending on clients-side uploaded features and logits \cite{he2020group,wu2023fedict,wu2022exploring}, which can be expressed as follows:
\begin{equation}
	\begin{array}{l}
	\mathop {\arg \min }\limits_{{W^S}} \sum\limits_{(X_i^k,y_i^k) \in {\mathcal{D}^k}} [{{L_{CE}}({\sigma _0}({f^S}(\underbrace {f_e^k(X_i^k)}_{{\rm{uploaded\, features}}})),y_i^k)} \\+ \lambda  \cdot {{KL}}({\sigma _0}({f^S}(\underbrace {f_e^k(X_i^k)}_{{\rm{uploaded\, features}}}))||{\sigma _1}(\underbrace {{f^k}(X_i^k)}_{{\rm{uploaded\, logits}}}))],
	\end{array}
\end{equation} 
where $\sigma_1(\cdot)$ is the transform mapping for local logits, and $\lambda$ is the distillation weighting factor. 
Contrastively, client $k$ performs local model parameters update with the server-side downloaded global logits, and optimizes the following loss function:
\begin{equation}
	\begin{array}{l}
	\mathop {\arg \min }\limits_{{W^k}} \sum\limits_{(X_i^k,y_i^k) \in {\mathcal{D}^k}} [{{L_{CE}}({\sigma _0}({f^k}(X_i^k)),y_i^k)} \\ + \mu  \cdot {{KL}}({\sigma _0}({f^k}(X_i^k))||{\sigma _2}(\underbrace {{f^S}(\underbrace {f_e^k(X_i^k)}_{{\rm{uploaded\, features}}})}_{{\rm{downloaded\, global\, logits}}}))],
	\end{array}
\end{equation}
where $\sigma_2(\cdot)$ is the transform mapping for global logits, and $\mu$ is the distillation weighting factor.
Although SLIA-FE allows for heterogeneous on-device models without parameters transmission, participants need to agree on the feature dimensionality.
Besides, since the feature dimensionality of high-resolution images and long sequential data is often high, the overhead of features transmission is still significant for devices. 
%In addition, features are vulnerable to inversion attacks, inevitably compromising user privacy.

\item
\textbf{SLIA with Public Dataset (SLIA-PD).} For SLIA-PD, client $k$ aims to approach the average logits of all clients on a given sample $(X^O_i,y^O_i)$ in the public dataset $\mathcal{D}^O$ \cite{itahara2021distill,li2019fedmd}, that is:
\begin{equation}
\begin{array}{*{20}{l}}
{\mathop {\arg \min }\limits_{{W^k}} \sum\limits_{(X_i^O,y_i^O) \in {{\cal D}^O}} {{L_{CE}}\Big({\sigma _0}({f^k}(X_i^O)),} }\\
\; \; \; \; \; \; \; \; \; \; \; \;
{{\sigma _0}(\frac{1}{K}\sum\limits_l {\frac{{{f^l}(X_i^O)}}{U}} )\Big)},
\end{array}
\end{equation}
where $U$ is a hyper-parameter that controls the distribution of ensembled logits.
We claim that SLIA-PD not only further relaxes the constraints on model architectures across clients, but also enables exchanges of only logits with minuscule sizes during training, resulting in significantly lower communication overhead compared to previously mentioned architectures.
However, SLIA-PD relies on a public dataset whose distribution should be close to private data on clients \cite{liu2022communication}. 
As it is unlikely to collect satisfactory public data without knowing data distribution of clients, this architecture is impractical in reality.
\end{itemize}

\begin{table*}[t]
	\caption{Comparison of FedCache with other PFL architectures in terms of model heterogeneity supportability, communication efficiency, dependency on public data, whether enable asynchronous optimization, and communication protocol.}
        \setlength{\tabcolsep}{6pt}
	\renewcommand\arraystretch{1.5}
	\centering
 \begin{adjustbox}{center}
	\begin{tabular}{l|c|c|c|c|c}
\hline
\multicolumn{1}{c|}{\textbf{Architecture}} & \textbf{\begin{tabular}[c]{@{}c@{}}Model Hetero.\\ Supportability\end{tabular}} & \textbf{\begin{tabular}[c]{@{}c@{}}Comm.\\ Efficiency\end{tabular}} & \textbf{\begin{tabular}[c]{@{}c@{}}No Dependency\\ on Public Data\end{tabular}} & \textbf{\begin{tabular}[c]{@{}c@{}}Asynchronous\\ Optimization\end{tabular}} & \textbf{\begin{tabular}[c]{@{}c@{}}Communication\\ Protocol\end{tabular}}     \\ \hline
PIA                                        & Partial Hetero.                                                                             & Low                                                                 & Yes                                                                            & No                                                                           & Model Parameters    \\
\hline
CLIA                                       & Complete Hetero.                                                                            & High                                                                & Yes                                                                            & No                                                                           & Class-grained Logits           \\
\hline
SLIA-FE                                    & \begin{tabular}[c]{@{}c@{}}Complete Hetero. with\\ Features Dim. Agreement\end{tabular}                                                                          & Medium                                                              & Yes                                                                            & Yes                                                                          & \begin{tabular}[c]{@{}c@{}}Sample-grained\\ Features and Logits\end{tabular}          \\
\hline
SLIA-PD                                    & Complete Hetero.                                                                            & High                                                                & No                                                                             & No                                                                           & Sample-grained Logits         \\ \hline
\textbf{FedCache}                            & \textbf{Complete Hetero.}                                                                   & \textbf{High}                                                       & \textbf{Yes}                                                                   & \textbf{Yes}                                                                 & \textbf{Sample-grained Logits} \\ \hline
\end{tabular}
\end{adjustbox}
	\label{cmp}
\end{table*}

\subsection{Motivation}
From the above analysis, we can conclude that existing PFL architectures cannot realize well-satisfied trade-offs among system performance, resource efficiency and without relying on public datasets, even if LIA gains advantages of remarkably reducing communication burden and tolerating heterogeneous models training over frequently-used PIA.
%以下部分是由于过度包装原因而删去的
% Compared to CLIA, SLIA can obtain superior precision but also induce client-side sacrifice in terms of resource overhead and public-data auxiliary.
% SLIA-FE allows the server to learn from clients' uploaded intermediate outputs and logits, extract new logits with the model on server, and communicate global logits back to clients.
% However, feature uploading increases non-negligible communication costs and easily occurs inversion assaults.
% Besides, SLIA-PD is dependent on a public dataset, and requires all clients to synchronize their input with public data to acquire average logits on the same publicly available sample for instructing clients-side local model training.
% On the contrary, CLIA dispenses with public datasets and feature communication, but only enables each client to learn per-class average logit vectors from other clients, with only labels being synchronized during the training process.
% Undoubtedly, the knowledge diversity learned in CLIA is limited by the number of classes, resulting in a substantial performance gap with previous architectures.
Motivated by the analysis above about PFL architectures, we attempt to answer the following question: \textcolor{black}{\textbf{how can a personalized federated learning architecture be designed to allow only logits transmission during the training process without the need for a public dataset, meanwhile significantly outperforming class-grained logits interaction-based architecture?} Concisely, our answer is to develop a knowledge cache-driven federated learning architecture with personalized distillation to optimize local models on clients.}
To optimize on-device models via knowledge distillation, we propose to keep a knowledge cache on the server, \textcolor{black}{which serves as the source of sample-grained knowledge for personalized distillation without public datasets.}
%which enables to fetch relevant personalized knowledge for each sample.
%and design a corresponding \textcolor{black}{knowledge filter mechanism} for each sample, through which related personalized knowledge can be fetched and ensemble distillation can be conducted on private samples of devices.
Specifically, the server-side knowledge cache keeps track of the latest knowledge of samples and leverages an information retrieval mechanism to search out the most relevant knowledge for each sample from cached knowledge.
%the knowledge cache needs to store the latest knowledge of samples, and match the most relevant knowledge for each sample through an information retrieval mechanism.}
%and integrate them according to knowledge relevance measured by hash distances between samples.
The searched knowledge from other clients is accompanied by reliable and effectual relevant representations, and is transferred to clients from which the sample originated for constructive distillation-based optimization.
On this basis, sample-grained logits interaction can be realized between the server and clients to ensure that on-device models learn sufficient personalized knowledge.

%allow the retrieval of knowledge relevant to the given sample.
%Specifically, we abandon the design of learning only $C$ distinct knowledge in CLIA, and instead allow each sample to learn the corresponding personalized knowledge.
%We also put aside the setting that keeps a model on the server in SLIA-FE since the server cannot extract global knowledge for a given sample without features transmission.
%Contrastively, we store the latest knowledge of samples on the server, matching the most relevant knowledge to each sample through an information retrieval mechanism, and integrating them on the server, with knowledge relevance measured by hash distances between samples.
%The integrated customized knowledge from other clients is accompanied by reliable and robust relevant representations, and is transferred to clients from which the sample originated for constructive optimization.

\begin{figure}[t]
	\centering
	\includegraphics[width=0.50\textwidth]{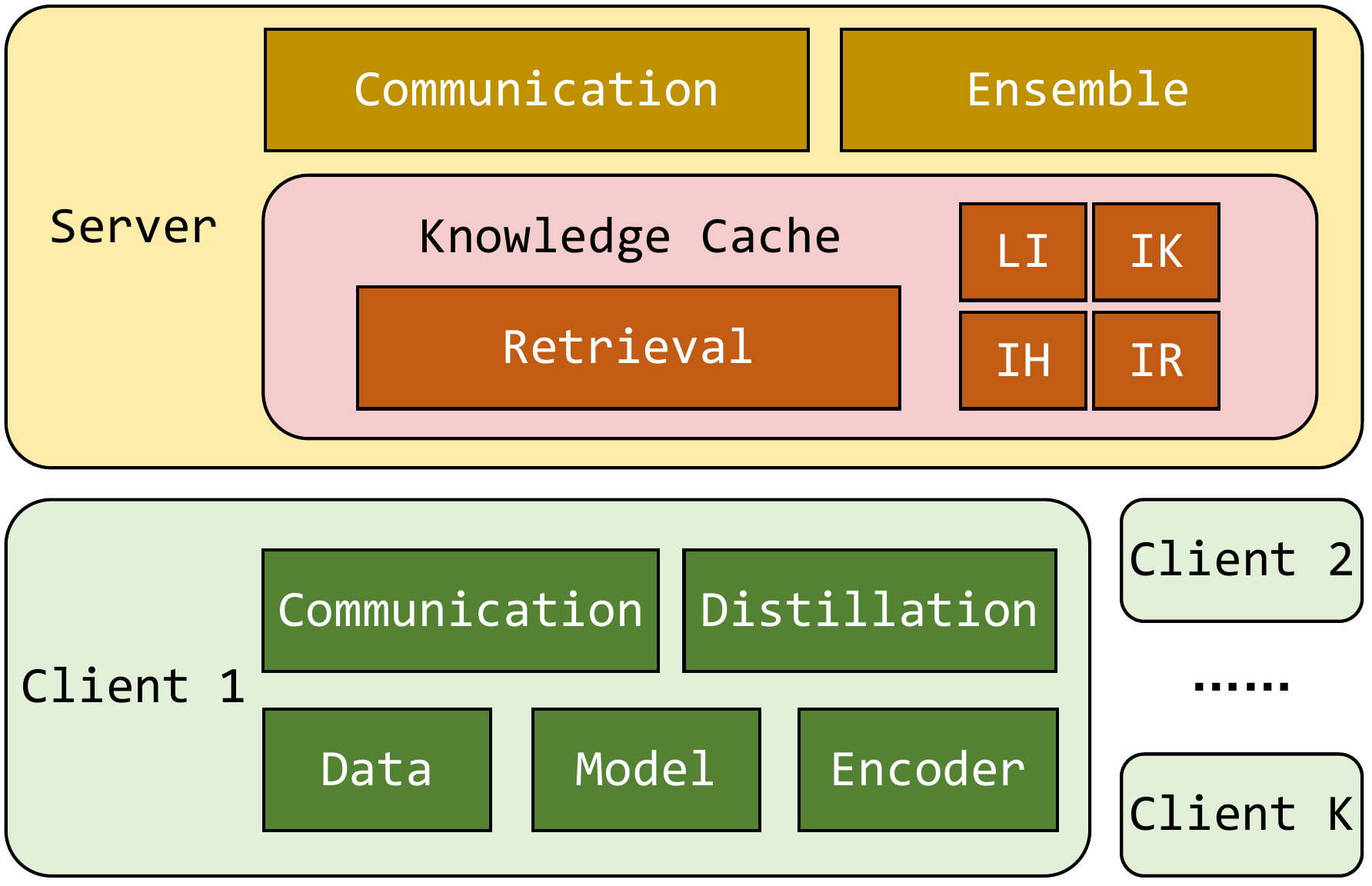}
	\caption{Functional module diagram of FedCache.}
	\label{modules}
\end{figure}

Based on the above insights, FedCache is proposed, whose comparisons with other FL architectures are shown in TABLE \ref{cmp}. 
%\textcolor{black}{Compared to existing architectures, FedCache enables collaborative training without model parameters aggregation, overcoming the restriction of requiring high-level model homogeneity among devices in PIA.} 
Compared to existing architectures, FedCache supports transferring sample-level logits without the assistance of public datasets during training, achieving superior performance compared to CLIA and overcoming the drawbacks of previous SLIA.
Besides, FedCache is a device-friendly architecture that enables complete model heterogeneity among clients, unlike other existing approaches either requiring partial model homogeneity or agreeing on the same feature dimension.
In addition, FedCache supports asynchronous interaction of logits required for PFL systems with devices of different capabilities, since it does not need to synchronously aggregate logits from different clients unlike previous methods \cite{jeong2018communication,itahara2021distill,li2019fedmd}.

%Compared to existing architectures, FedCache accommodates high-level model heterogeneity without public datasets by just transferring logits during training.
%In addition, FedCache supports asynchronous training and avoids features transmission, making it resistant to inversion attacks.
%In addition, clients can acquire sample-level global logits in FedCache, which improves auxiliary information and can achieve superior performance than CLIA.

\begin{figure}[t]
	\centering
	\includegraphics[width=0.5\textwidth]{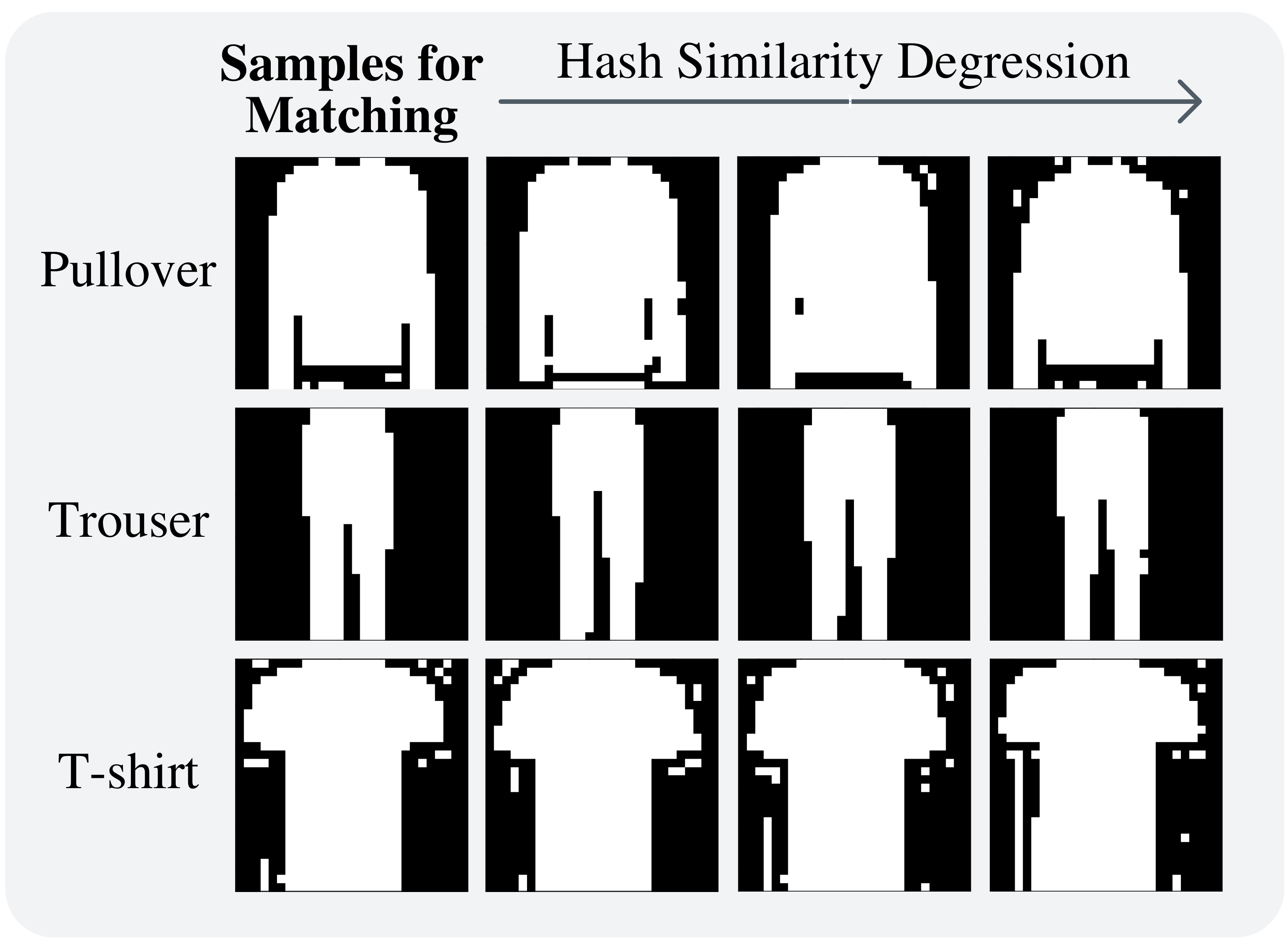}
	\caption{Sample matching results on FashionMNIST dataset with $R=3$.}
	\label{sim}
\end{figure}

\section{Knowledge Cache-driven Personalized Federated Learning}
\subsection{System Design}
The functional module diagram of FedCache is displayed in Fig. \ref{modules}, which consists of a server with three functional modules: (server-client) communication, (knowledge) ensemble, knowledge cache; and $K$ clients with five functional modules: (client-server) communication, (knowledge) distillation, data, model, and (sample) encoder.
Specifically, the ensemble module combines the fetched knowledge from the knowledge cache to obtain personalized knowledge to be distilled over clients;
the knowledge cache module is our designed self-organizing knowledge storage structure that facilitates fetching each client's relevant knowledge on the server side; 
the model module extracts knowledge from local data, and conducts model updates under the guidance of the distillation module;
in addition, the encoder module encodes private data into hash values to initialize the knowledge cache. \textcolor{black}{The encoder should be efficient, robust, and discriminative, ensuring that the hash values of local samples can be computed quickly and reliably reflect the semantic similarity among samples.}

During the initialization phase, the generated hash codes on clients are uploaded to the server in a single pass.
Then, HNSW \cite{malkov2018efficient} is performed in the server-side knowledge cache, aiming to retrieve $R$ most relevant samples for matching each sample measured by cosine similarity of hash values.
Fig. \ref{sim} displays the sample matching results on FashionMNIST \cite{xiao2017fashion} dataset.
As shown, the matched samples are very similar to the original sample, making the knowledge extracted from them beneficial to client-side distillation on the original sample.
During the training phase, each logits and index of private samples are uploaded to the server in each communication round. 
Then, $R$ best-matching knowledge with the highest hash similarity in the knowledge cache for each sample is fetched based on the pre-established similarity relations, followed by knowledge ensemble and then knowledge communication to corresponding clients for local distillation.
As the distillation phase only relies on highly relevant knowledge of clients' respective private data, the resulting model is locally adaptable and powerful for personalization tasks.
We will introduce the key procedures of FedCache in the subsequent subsections.

%As shown in Figure \ref{arch}, the hash code of samples encoded by a pre-trained deep neural network is uploaded to the server and then stored in the knowledge cache. 
%Next, the similarity relations between samples on clients are established during the graph building process of the HNSW algorithm \cite{malkov2018efficient}. 
%When training starts, all clients iteratively uploads logits (knowledge) extracted from local samples to the server, and the server retrieves for $R$ most matching knowledge with the maximum hash similarity based on the previously constructed similarity relations. The matched ensemble knowledge is averaged on the server and subsequently downloaded to corresponding clients for local distillation.
%We will detailly introduce each arithmetic of our proposed FedRC in the following subsections.

% aiming to retrieve relevant knowledge for each client with a controlled computational complexity without requiring synchronization of clients. 
% Specifically, the operation of the knowledge cache is of two main phases: initialization and training.

\subsection{Knowledge Cache}
The knowledge cache on the server is proposed to asynchronously fetch relevant knowledge for an arbitrary local sample with controllable computation complexity, where the corresponding hash values of samples from which relevant knowledge is extracted should be one of the $R$-nearest neighbors of the hash value of the original sample.
%As directly matching knowledge with the closest hash values requires frequent client-side real-time interactions with the server, which imposes a huge synchronization overhead and is vulnerable to client-side offline scenarios.
Guided by the above design, we preserve multiple pairs in the knowledge cache, including label-to-index pairs ($LI$), index-to-knowledge pairs ($IK$), index-to-hash pairs ($IH$), and index relations pairs ($IR$), where each pair enables mapping the first element to the second element. 
On this basis, the knowledge cache is of two main phases: initialization and training.
%the operation of the knowledge cache is of two main phases: initialization and training.

\color{black}
The initialization process includes the following steps:
\begin{itemize}
	\item
	\textbf{Pairs initialization.}
	The uploaded hash value $h^k_i$ corresponding to each sample index $(k,i)$ is stored in $IH$. In addition, indexes are added to $LI$ according to their corresponding label classes, and the knowledge corresponding to each given index is initialized to zeros in $IK$, i.e.
	\begin{equation}
		IH(k,i) \leftarrow h_i^k,
		\label{init1}
	\end{equation}
	\begin{equation}
		  LI(y_i^k)\leftarrow LI(y_i^k)  \cup \{(k,i)\},
		\label{init2}
	\end{equation}
	\begin{equation}
		IK(k,i) \leftarrow (\underbrace {0,...,0}_{C{\rm{ }}zeros}).
		\label{init3}
	\end{equation}
        As $LI$ only allows relations to be built within the sample index range of the same label class, it is expected that the number of candidate samples used for matching will be reduced, improving the computation efficiency of the relations establishment in the following step.
	\item
	\textbf{Build relations.}
    For each given sample index $(k,i)$, we relate it to $R$ indexes $\{(l_1,j_1),(l_2,j_2),...,(l_R,j_R)\}$ whose hash values have the greatest cosine similarity to the hash value of the given sample among all the candidate hashes, i.e.
	\begin{equation}
		\begin{array}{l}
\mathop {\arg \max }\limits_{{{(l_1,j_1)}},{{(l_2,j_2)}},...,{{(l_R,j_R)}}} \sum\limits_{m = 1}^R {\cos (IH(k,i),IH({l_m},{j_m}))} ,\\
s.t.
\left\{ \begin{array}{l}
{l_{{n_1}}} \ne {l_{{n_2}}} \vee {j_{{n_1}}} \ne {j_{{n_2}}},\forall {n_1},{n_2} \wedge {n_1} \ne {n_2},\\
(k,i) \in LI({y^*}) \wedge ({l_m},{j_m}) \in LI({y^*}),\exists {y^*},\\
%{(hr_i^k)_m} = IH({l_m},{j_m}),\forall m\\
{n_1},{n_2},m \in \{ 1,2,...,R\} ,\\
{y^*} \in \{ 1,2,...,C\}, 
\end{array} \right.
\label{hnsw}
\end{array}
		\label{rel1}
	\end{equation}
        during which HNSW \cite{malkov2018efficient} is adopted to achieve the $R$-nearest neighbors retrieval.
        Then, the retrieved results related to each sample index are saved in $IR$ for subsequent access, i.e.
	\begin{equation}
		IR(k,i)\leftarrow\{{{{(l_1,j_1)}},{{(l_2,j_2)}},...,{{(l_R,j_R)}}}\}
        %IR(k,i)\leftarrow\{(hr_i^k)_1,{{(hr_i^k)}_2},...,(hr_i^k)_R\}.
		\label{rel2}
	\end{equation}

\end{itemize}

%改到此处，最后要补一句hash和knowledge怎么来的自后续章节中阐述
During the training process, the following steps should be performed for each given sample index:
\begin{itemize}
    \item 
    \textbf{Knowledge fetching.} The most relevant knowledge can be fetched in the knowledge cache $KC$ based on a provided sample index: for a newly uploaded sample index $(k,i)$, the corresponding knowledge is obtained and returned according to 1) $IR$ which stores relevant sample indexes of $(k,i)$, and 2) $IK$ which transforms relevant indexes to knowledge, that is:
\begin{equation}
	KC(h_i^k;k,i) = IK(IR(k,i)).
	\label{fetch2}
\end{equation}
As knowledge fetching requires only the clients requesting knowledge to be online, clients can asynchronously perform fetched knowledge-based optimization.
    \item
    \textbf{Knowledge update.} $IK(k,i)$ is updated with the knowledge $z_i^k$ corresponding to the given sample index $(k,i)$, so that the latest knowledge can be fetched on the next access, i.e.
\begin{equation}
    IK(k,i) \leftarrow z_i^k.
    \label{update}
\end{equation}
\end{itemize}
%\textcolor{black}{The ways that the knowledge cache interacts with clients and other functional modules on the server are described in the following subsections.}

\begin{figure*}[t]
	\centering
	\includegraphics[width=1.0\textwidth]{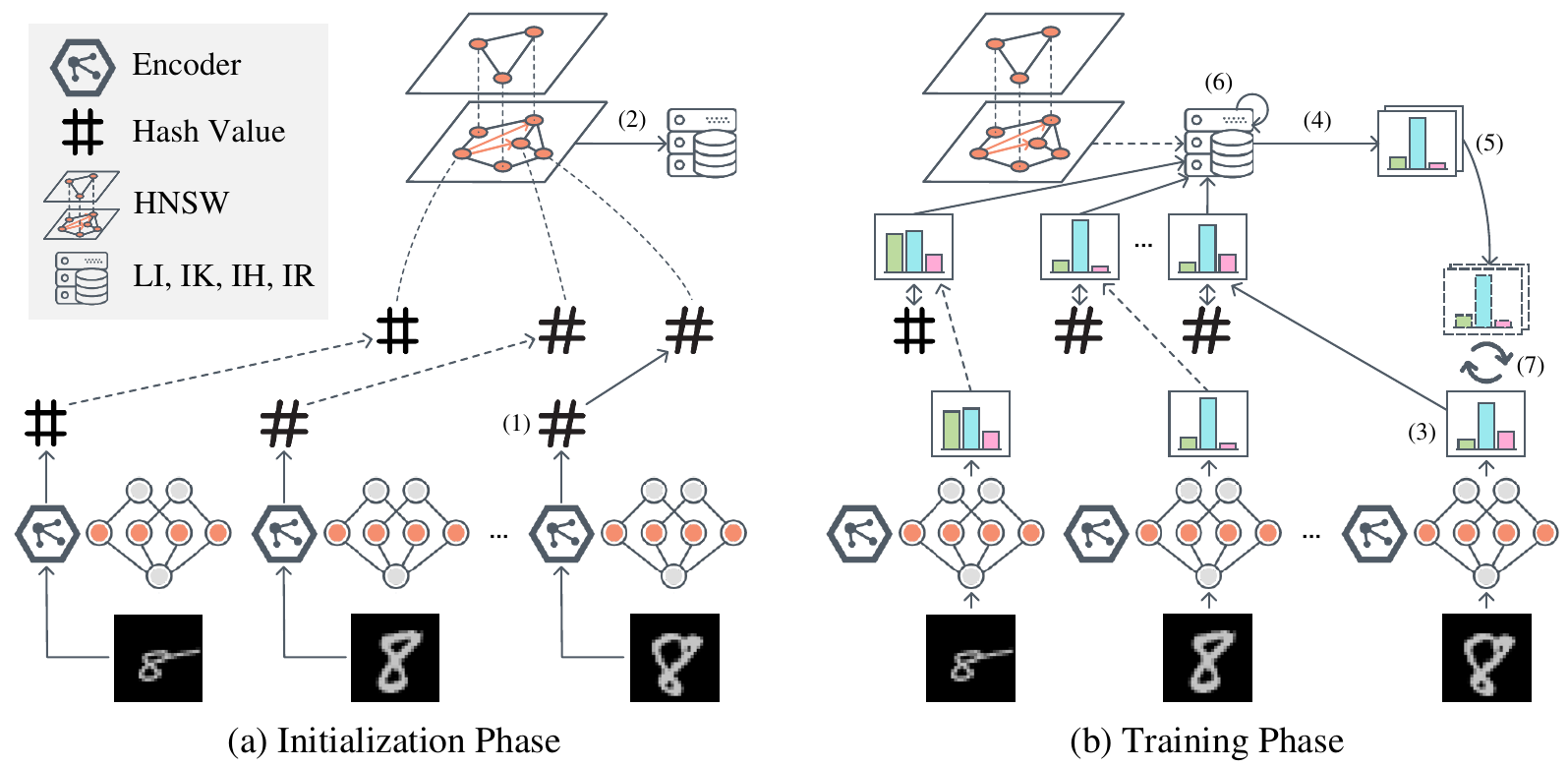}
	\caption{Overview of executing procedure of FedCache. (1) Hash encoding and uploading. (2) Knowledge cache initialization. (3) Knowledge extraction and uploading. (4) Knowledge fetching. (5) Knowledge ensemble and distributing. 
 (6) Knowledge update. (7) Knowledge acceptance and distillation.}
	\label{arch}
\end{figure*}

\subsection{Knowledge Cache-driven Personalized Distillation}
%As a communication-efficient FL paradigm, federated distillation does not exchange model parameters during the training process.
%However, most existing federated distillation methods \cite{li2019fedmd,itahara2021distill} unrealistically rely on a public dataset whose distribution is similar to the clients' private data \cite{liu2022communication}, where the selected clients learn their average logits extracted based on a public dataset.
We optimize on-device models with personalized federated distillation, where knowledge of samples similar to each client's private data is fetched from the knowledge cache.
On this basis, each client performs ensemble distillation on fetched knowledge for constructive optimization of on-device models.
Specifically, a pre-trained deep neural network $f^h(\cdot)$ is adopted as an encoder to generate the hash values of samples on clients during initialization, i.e.
\begin{equation}
    \, h_i^k = {f^h}(X_i^k),
    \label{hash}
\end{equation}
and such hash values with corresponding sample indexes and labels are uploaded to the server for initializing the knowledge cache according to Eq. (\ref{init1}, \ref{init2}, \ref{init3}, \ref{rel1},\ref{rel2}).

During training on each given sample $(X_i^k,y_i^k)$, client $k$ first extracts knowledge $z_i^k$ on $X_i^k$, and then uploads $z_i^k$ with corresponding sample index $(k,i)$ to the server, where:
\begin{equation}
z_i^k = {f^k}(X_i^k).
\end{equation}
Then, the $R$ knowledge related to sample index $(k,i)$ is fetched from the knowledge cache $KC$ according to Eq. (\ref{fetch2}), that is,
\begin{equation}
		{(zr_i^k)_1},{(zr_i^k)_2},...,{(zr_i^k)_R} = KC(h_i^k;k,i),	
	\label{fetch1}
\end{equation}
%where $KC$ is the knowledge cache, whose results is obtained from Eq. (\ref{fetch2}), 
where $(zr_i^k)_s$ is the $s$-th knowledge fetched for the given sample index $(k,i)$. The fetched knowledge is ensembled in an average manner, which can be expressed as:
\begin{equation}
	\overline {zr} _i^k = \frac{1}{R}\sum\limits_{s = 1}^R {{{(zr_i^k)}_s}}.
	\label{avg}
\end{equation}
Subsequently, the ensembled knowledge is distributed to client $k$ for performing distillation-based local model optimization weighted by factor $\beta$, which is defined as follows:
\begin{equation}
	\begin{array}{l}
\; \; \; \; \mathop {\arg \min }\limits_{{W^k}} {J^k}({W^k}) \\
= \mathop {\arg \min }\limits_{{W^k}} \sum\limits_{(X_i^k,y_i^k) \in {{\cal D}^k}} {[{L_{CE}}(\tau ({f^k}(X_i^k)),y_i^k)} \\
 \; \; \; \; \; \; \; \; \; \; \; \;+ \beta  \cdot KL(\tau ({f^k}(X_i^k))||\tau (\overline {zr} _i^k))].
\end{array}
	\label{j-client}
\end{equation}
\color{black}

\subsection{Formal Description of FedCache}
The overview of executing procedure of FedCache is shown in Fig. \ref{arch}, and the execution processes of FedCache on client $k$ and the server are respectively formulated in Algorithms \ref{alg1} and \ref{alg2}. 
From the overall perspective, we allow personalized local models on devices to distill ensembled knowledge on the samples similar to private data with the assistance of the server-side knowledge cache.

Specifically, FedCache consists of the following steps:
\begin{itemize}
    \item 
\textbf{Hash Encoding and Uploading.} For each sample from a given client, a hash value is encoded based on the pre-trained local encoder according to Eq. (\ref{hash}), (Algorithm \ref{alg1}, line 3). This hash value is uploaded to the server along with the corresponding label and sample index (Algorithm \ref{alg1}, line 4).
As the encoder is a deep pre-trained neural network with a large number of superimposed non-linear mapping and the dimensionality of the output code is much smaller than that of data, sharing hash values with the server is privacy-preserving.
    \item
\textbf{Knowledge Cache Initialization.} The server accepts the uploaded information from clients (Algorithm \ref{alg2}, line 3) and establishes relations between sample indexes in the knowledge cache according to Eq. (\ref{init1}, \ref{init2}, \ref{init3}, \ref{rel1}, \ref{rel2}), such that each sample can be indexed to $R$-related samples (Algorithm \ref{alg2}, line 4-6).
    \item
\textbf{Knowledge Extraction and Uploading.} Clients extract logits (Algorithm \ref{alg1}, line 9) and upload logits with corresponding sample indexes to the server (Algorithm \ref{alg1}, line 10). This step is an alternative to the parameters/features uploading step of PIA and SLIA-FE. As the size of the logits and sample indexes are several orders of magnitude smaller than that of the model parameters or features, the communication burden can be significantly reduced.
    \item
\textbf{Knowledge Fetching.} The server accepts the sample indexes uploaded by the clients (Algorithm \ref{alg2}, line 10), and fetches $R$-nearest matching knowledge from the knowledge cache based on pre-established sample index relations (Algorithm \ref{alg2}, line 11).
\textcolor{black}{This step enables on-device models to obtain sample-level granularity of knowledge without being limited by the number of classes.}
%This step allows FedCache for sample-level granularity of knowledge without being limited by the number of classes.
    \item
\textbf{Knowledge Ensemble and Distributing.}
    The fetched knowledge is ensembled on the server according to Eq. (\ref{avg}) (Algorithm \ref{alg2}, line 12), and is subsequently distributed to corresponding clients (Algorithm \ref{alg2}, line 13). This step is also communication-efficient since only logits are transferred between the server and clients.
    \item 
\textbf{Knowledge Update.} The stored knowledge in the knowledge cache is updated based on the newly-uploaded knowledge (in Algorithm \ref{alg2}, line 10) according to Eq. (\ref{update}). (Algorithm \ref{alg2}, line 14)
    \item
\textbf{Knowledge Acceptance and Distillation.} The clients receive the ensembled knowledge distributed from the server (Algorithm \ref{alg1}, line 11) and optimize client-side local models according to Eq. (\ref{j-client}) (Algorithm \ref{alg1}, lines 12-13). \textcolor{black}{This step can be performed asynchronously on each client without waiting for other clients to finish their previous steps.}
%clients learn personalized knowledge extracted from related samples via only transferring logits.
\end{itemize}

\begin{algorithm}[t]
	\caption{FedCache on Client $k$.}
	\SetAlgoLined
        \LinesNumbered
       
        \label{alg1}
         \begin{spacing}{1.1}
	%\KwIn{$\mathcal{D}^k$, $M^K$, $f^{h}$}%输入参数
	%\KwOut{Trained $W^k$}%输出
        %\begin{algorithmic}[1]
	//Initialization process\\
	\ForEach{$(X^k_i,y^k_i) \in \mathcal{D}^k$}
	{
		 $h_i^k\leftarrow f^h(X_i^k)$\\
	   Upload $h_i^k$ with index $(k,i)$ and label $y^k_i$ to the server 
	}
	//Training process\\
	\Repeat{Training stop}
	{
		\ForEach{$(X^k_i,y^k_i) \in \mathcal{D}^k$}
		{
			$z_i^k \leftarrow {f^k}(X_i^k)$\\
			Upload $z_i^k$ with index $(k,i)$ to the server\\
			Download averaged ensemble knowledge $\overline {zr} _i^k$ from the server\\
			${W^k} \leftarrow {W^k} - lr \cdot {\nabla _{{W^k}}}J^k({W^k})$ \\ \Comment{Optimize Eq. (\ref{j-client})}\\
		}
	}
    \end{spacing}
\end{algorithm}

\begin{algorithm}[t]
        \newcommand{\removelatexerror}{\let\@latex@error\@gobble}
	\caption{FedCache on the Server.}
	\SetAlgoLined
        \LinesNumbered
	\label{alg2}
	%\KwIn{None}%输入参数
	%\KwOut{None}%输出
        \begin{spacing}{1.1}
	//Initialization process\\
	\Repeat{Receive all indexes $(k,i)$ from $K$ clients}
	{
		Receive $h_i^k$ with index $(k,i)$ and label $y_i^k$ from client $k$\\
		Update $LI$, $IK$, $IH$ according to Eq. (\ref{init1}, \ref{init2}, \ref{init3})
	}
	Build relations via HNSW \cite{malkov2018efficient} according to Eq. (\ref{rel1}) and Eq. (\ref{rel2})
	\\
	//Training process\\
	\Repeat{Training stop}
	{
		\ForEach{$(k,i)$}
		{
			Receive $(k,i)$ and $z_i^k$ from client $k$\\
			Fetch $R$ related knowledge from the knowledge cache according to Eq. (\ref{fetch1}) and Eq. (\ref{fetch2})\\
			Obtain ensembled fetched knowledge $\overline {zr} _i^k$ according to Eq. (\ref{avg})\\
			Send $\overline {zr} _i^k$ to client $k$\\
            Update knowledge cache according to Eq. (\ref{update})
			%		Receive $\{z_1^k,z_2^k,...,z_{{N^k}}^k\}$ with indexes $\{(k,1), (k,2),...,(k,N^k)\}$ from client $k$\\
			%		Retrieve ensemble knowledge $\{\{(zr_1^k)_1,...,(zr_1^k)_R\},...,\{(zr_{N^k}^k)_1,...,(zr_{N^k}^k)_R\}\}$ corresponding to $R$ nearest hash values, according to \cite{malkov2018efficient}
			%知识平均
		}
	}
\end{spacing}
\end{algorithm}

\section{Experiments}
\label{exper}
\subsection{Experimental Setup}
\label{exp-set}
\subsubsection{Datasets and Preprocessing}
We conduct experiments on four common datasets, MNIST \cite{lecun1998gradient}, FashionMNIST \cite{xiao2017fashion}, CIFAR-10 \cite{krizhevsky2009learning} and CINIC-10 \cite{darlow2018cinic}. 
\textcolor{black}{Following \cite{wu2023fedict}, we adopt the data partitioning scheme in FedML \cite{he2020fedml}, which uses a hyperparameter $\alpha$ ($\alpha > 0$) to control the degree of local data distribution differentiation among devices. As $\alpha$ decreases, the data distributions among devices show greater degrees of heterogeneity. To evaluate FedCache on personalized data, we use the same data partitioning strategy for both the complete training and testing datasets, ensuring that the label distributions of training and testing local data are consistent on each device.}
In all of our main experiments, we partition each dataset into 300 non-independent identically distributed copies for training and testing on $K=300$ different clients, and the hyper-parameter $\alpha$ is set to 1.0. Each client runs locally for one epoch before model aggregation or feature/knowledge transfer.

\subsubsection{Benchmarks and Criteria}
\label{criterions}
To fully demonstrate the effectiveness of FedCache, we compare it with the state-of-art PFL methods with various architectures, including FMTL \cite{mills2021multi} and pFedMe \cite{t2020personalized} based on PIA, FedDKC \cite{wu2022exploring} and FedICT \cite{wu2023fedict} based on SLIA-FE, and FD \cite{jeong2018communication} based on CLIA. Among all the architectures, SLIA-PD is discarded because of its impractical reliance on public datasets.
%Since SLIA-PD impractically relies on a public dataset during training, 
%we compare FedCache with existing PFL methods that do not require public datasets, 
The precision of benchmark algorithms is measured by Maximum Average User model Accuracy \cite{mills2021multi} (MAUA). 
Moreover, we denote the communication overhead to reach a given average UA $acc$ as $acc@$, measuring system communication efficiency with different $acc@$ according to the actual system performance, as shown in TABLE \ref{acc-table}. 
%改到此处
% \textcolor{black}{The speed-up ratios of individual methods are obtained by calculating the ratio of the communication efficiency to the method with the highest communication overhead among all benchmark algorithms under the same experimental setting.}
We also calculate the speed-up ratio of each method by comparing the ratio of communication overhead between the one with the highest communication overhead of all benchmark algorithms and this method under the same experimental settings.
In addition, our MAUA results are obtained in a reasonable training time, when the algorithm reaches convergence or the total communication overhead reaches the given limitation, such as 55G and 19G for CIFAR-10 and CINIC-10 datasets, respectively.

\begin{table}[t!]
\renewcommand\arraystretch{1.2}
	\caption{The $acc@$ used in different experiments to measure system communication overhead.}
 \centering
\begin{tabular}{c|cccc}
\hline
                                                                                   & \multicolumn{4}{c}{\textbf{Dataset}}                                                                                                           \\
\multirow{-2}{*}{\textbf{\begin{tabular}[c]{@{}c@{}}Model\\ Setting\end{tabular}}} & \textbf{MNIST}                      & \begin{tabular}[c]{@{}c@{}}\textbf{Fashion}\\ \textbf{MNIST}\end{tabular} & \textbf{CIFAR-10}                   & \textbf{CINIC-10}                   \\ \hline
\textbf{\begin{tabular}[c]{@{}c@{}}Model\\ Homo.\end{tabular}}                     & {87@} & {77@}                              & {43@} & {40@} \\ \hline
\textbf{\begin{tabular}[c]{@{}c@{}}Model \\ Hetero.\end{tabular}}                  & {83@} & {77@}                              & {41@} & {41@} \\ \hline
\end{tabular}
\label{acc-table}
\end{table}

%\vspace {0.8cm}

\subsubsection{Models}
For the deep pre-trained encoder, we adopt MobileNetV3 \cite{howard2019searching} pre-trained on ImageNet \cite{deng2009imagenet}, with the last fully connected layer removed.
In addition, we consider 4 different model architectures, where $\{A^C_1,A^C_2,A^C_3\}$ are for clients, and $A^S$ is for the server, and the main configurations of four adopted models are shown in TABLE \ref{model-conf}.
It is worth noting that the model on the server does not contain the foremost Conv+Batch+ReLU layers to fit the training requirements of \cite{wu2022exploring,wu2023survey}.
Moreover, both client-side model homogeneity and heterogeneity are considered in our experiments.
Specifically, for the experiments with homogeneous models, we compare FedCache with all aforementioned benchmark algorithms, and all clients adopt the model architecture $A^C_3$. For the experiments with heterogeneous models, FedCache only compares with the benchmarks that support model heterogeneity among clients, including FedDKC, FedICT and FD, and clients with residuals of index mod 3 of 0, 1 and 2 are assigned with model architectures $A^C_1$, $A^C_2$ and $A^C_3$ respectively.
%For transportation mode detection, we compare FedCache with all aforementioned benchmarks with the following model architecture setting: we adopt heterogeneous client models on FedCache, FedDKC, FedICT and FD, and clients with index mod 3 of 0, 1, and 2 are assigned to model architectures $A^C_4$, $A^C_5$ and $A^C_6$, respectively. In addition, we conduct three groups of experiments on FMTL and pFedMe with homogeneous models, where all clients are assigned with model architectures $A^C_4$, $A^C_5$ and $A^C_6$, respectively.

%\subsubsection{Encoder}
 
%For transportation mode detection, we apply random orthogonal matrix transformations to normalized sensor data to retain cosine similarity between samples.

\subsubsection{Hyper-parameter Settings}
We adopt stochastic gradient descent with a learning rate $lr=0.01$ and a batch size of 8 in all the experiments.
In addition, the hyper-parameters of benchmark algorithms are set as follows:
\begin{itemize}
    \item
    For pFedMe, we set $\eta=0.005$, $\lambda=15$ and $\beta=1$ according to \cite{t2020personalized}.
    \item
    For MTFL, we adopt the FedAvg optimization strategy \cite{mills2021multi}, with other hyper-parameters following the default setting in \cite{mtflurl}.
    \item
    For FedDKC, we adopt KKR as the knowledge refinement strategy, with $\beta=1.5$ and $T=0.12$ according to \cite{wu2022exploring}.
    \item
    For FedICT, we adopt the similarity-based LKA strategy, with $\beta=\lambda=\mu=1.5$ and $T=3.0$ according to \cite{wu2023fedict}.
    \item
    For FD, no individualized hyper-parameters are required \cite{jeong2018communication}.
\end{itemize}
Finally, for our proposed FedCache, we set $\beta=1.5$ and $R = 16$. The impact of hyper-parameters on system performance will be investigated in the ablation study.

\begin{table}[t]
\renewcommand\arraystretch{1.2}
	\caption{Main configurations of four adopted models. The height and width of the input images are noted as $H$ and $W$, respectively.}
 \centering
\begin{tabular}{l|c|c|c}
\hline
\multicolumn{1}{c|}{\textbf{Model}} & \textbf{Notation} & \textbf{Feat. Shape}    & \textbf{Params} \\ \hline
ResNet-small                            & $A^C_1$                & \multirow{4}{*}{$H\times W \times 16$} & 76.2K           \\
ResNet-medium                           & $A^C_2$                &                         & 171.2K          \\
ResNet-large                           & $A^C_3$                &                         & 266.1K          \\
ResNet-server                           & $A^S$               &                         & 588.2K          \\ \hline
\end{tabular}
\label{model-conf}
\end{table}

\begin{table*}[!t]
	\caption{MAUA (\%), communication overhead and communication efficiency speed-up ratio on homogeneous on-device models. Some methods are unable to calculate the communication overhead with corresponding speed-up ratios as they cannot achieve the MAUA in TABLE \ref{acc-table} under given experimental settings, and their corresponding items are denoted by -. The same as below.}
	\centering
	\setlength{\tabcolsep}{8pt}
	\renewcommand\arraystretch{1.15}
\begin{tabular}{c|l|cc|ccc}
\hline
                                        & \multicolumn{1}{c|}{}                                  & \multicolumn{2}{c|}{\textbf{Model}}                               & \multicolumn{3}{c}{\textbf{Metric}}                                                                             \\
\multirow{-2}{*}{\textbf{Dataset}}      & \multicolumn{1}{c|}{\multirow{-2}{*}{\textbf{Method}}} & \textbf{Client}                          & \textbf{Server}       & \textbf{MAUA (\%)} & { \textbf{Comm. (G)}} & { \textbf{Speed-up Ratio}} \\ \hline
                                        & pFedMe                                                 & \multicolumn{1}{c|}{}                     &                       & 94.89              & 13.25                                     & $\times$1.0                                           \\
                                        & MTFL                                                   & \multicolumn{1}{c|}{}                     & \multirow{-2}{*}{$A^C_3$}  & 95.59              & 7.77                                      & $\times$1.7                                           \\ \cline{4-4}
                                        & FedDKC                                                 & \multicolumn{1}{c|}{}                     &                       & 89.62              & 9.13                                      & $\times$1.5                                           \\
                                        & FedICT                                                 & \multicolumn{1}{c|}{}                     & \multirow{-2}{*}{$A^S$} & 84.62              & -                                         & -                                              \\ \cline{4-4}
                                        & FD                                                     & \multicolumn{1}{c|}{}                     &                       & 84.19              & -                                         & -                                              \\
\multirow{-6}{*}{\textbf{MNIST}}        & \textbf{FedCache}                                        & \multicolumn{1}{c|}{\multirow{-6}{*}{$A^C_3$}} & \multirow{-2}{*}{-}   & 87.77              & \textbf{0.99}                             & \textbf{$\times$13.4}                                 \\ \hline
                                        & pFedMe                                                 & \multicolumn{1}{c|}{}                     &                       & 81.57              & 20.71                                     & $\times$1.0                                           \\
                                        & MTFL                                                   & \multicolumn{1}{c|}{}                     & \multirow{-2}{*}{$A^C_3$}  & 83.92              & 12.33                                     & $\times$1.7                                           \\ \cline{4-4}
                                        & FedDKC                                                 & \multicolumn{1}{c|}{}                     &                       & 78.24              & 8.43                                      & $\times$2.5                                           \\
                                        & FedICT                                                 & \multicolumn{1}{c|}{}                     & \multirow{-2}{*}{$A^S$} & 76.90              & 13.34                                     & $\times$1.6                                           \\ \cline{4-4}
                                        & FD                                                     & \multicolumn{1}{c|}{}                     &                       & 76.32              & -                                         & -                                              \\
\multirow{-6}{*}{\textbf{FashionMNIST}} & \textbf{FedCache}                                        & \multicolumn{1}{c|}{\multirow{-6}{*}{$A^C_3$}} & \multirow{-2}{*}{-}   & 77.71              & \textbf{0.08}                             & \textbf{$\times$258.9}                                \\ \hline
                                        & pFedMe                                                 & \multicolumn{1}{c|}{}                     &                       & 37.49              & -                                         & -                                              \\
                                        & MTFL                                                   & \multicolumn{1}{c|}{}                     & \multirow{-2}{*}{$A^C_3$}  & 43.43              & 52.99                                     & $\times$1.0                                           \\ \cline{4-4}
                                        & FedDKC                                                 & \multicolumn{1}{c|}{}                     &                       & 45.87              & 11.46                                     & $\times$4.6                                           \\
                                        & FedICT                                                 & \multicolumn{1}{c|}{}                     & \multirow{-2}{*}{$A^S$} & 43.61              & 10.69                                     & $\times$5.0                                           \\ \cline{4-4}
                                        & FD                                                     & \multicolumn{1}{c|}{}                     &                       & 42.77              & -                                         & -                                              \\
\multirow{-6}{*}{\textbf{CIFAR-10}}     & \textbf{FedCache}                                        & \multicolumn{1}{c|}{\multirow{-6}{*}{$A^C_3$}} & \multirow{-2}{*}{-}   & 44.42              & \textbf{0.19}                             & \textbf{$\times$278.9}                                \\ \hline
                                        & pFedMe                                                 & \multicolumn{1}{c|}{}                     &                       & 31.65              & -                                         & -                                              \\
                                        & MTFL                                                   & \multicolumn{1}{c|}{}                     & \multirow{-2}{*}{$A^C_3$}  & 34.09              & -                                         & -                                              \\ \cline{4-4}
                                        & FedDKC                                                 & \multicolumn{1}{c|}{}                     &                       & 43.95              & 4.12                                      & $\times$1.3                                           \\
                                        & FedICT                                                 & \multicolumn{1}{c|}{}                     & \multirow{-2}{*}{$A^S$} & 42.79              & 5.50                                      & $\times$1.0                                           \\ \cline{4-4}
                                        & FD                                                     & \multicolumn{1}{c|}{}                     &                       & 39.36              & -                                         & -                                              \\
\multirow{-6}{*}{\textbf{CINIC-10}}     & \textbf{FedCache}                                        & \multicolumn{1}{c|}{\multirow{-6}{*}{$A^C_3$}} & \multirow{-2}{*}{-}   & 40.45              & \textbf{0.07}                             & \textbf{$\times$78.6}                                 \\ \hline
\end{tabular}
\label{model-homo}
%\end{table*}
\vspace {0.6cm}

%\begin{table*}
\caption{MAUA (\%), communication overhead and communication efficiency speed-up ratio on heterogeneous on-device models.}
	\centering
	\setlength{\tabcolsep}{8pt}
	\renewcommand\arraystretch{1.15}
\begin{tabular}{c|l|cc|ccc}
\hline
                                        & \multicolumn{1}{c|}{}                                  & \multicolumn{2}{c|}{\textbf{Model}}                                     & \multicolumn{3}{c}{\textbf{Metric}}                                                                             \\
\multirow{-2}{*}{\textbf{Dataset}}      & \multicolumn{1}{c|}{\multirow{-2}{*}{\textbf{Method}}} & \textbf{Client}                                & \textbf{Server}       & \textbf{MAUA (\%)} & {\textbf{Comm. (G)}} & { \textbf{Speed-up Ratio}} \\ \hline
                                        & FedDKC                                                 & \multicolumn{1}{c|}{}                           &                       & 85.38              & 10.53                                     & $\times$1.0                                           \\
                                        & FedICT                                                 & \multicolumn{1}{c|}{}                           & \multirow{-2}{*}{$A^S$} & 80.53              & -                                         & -                                              \\ \cline{4-4}
                                        & FD                                                     & \multicolumn{1}{c|}{}                           &                       & 79.90               & -                                         & -                                              \\
\multirow{-4}{*}{\textbf{MNIST}}        & \textbf{FedCache}                                        & \multicolumn{1}{c|}{\multirow{-4}{*}{$A^C_1$, $A^C_2$, $A^C_3$}} & \multirow{-2}{*}{-}   & 83.94              & \textbf{0.10}                              & \textbf{$\times$105.3}                                \\ \hline
                                        & FedDKC                                                 & \multicolumn{1}{c|}{}                           &                       & 77.96              & 12.64                                     & $\times$1.0                                           \\
                                        & FedICT                                                 & \multicolumn{1}{c|}{}                           & \multirow{-2}{*}{$A^S$} & 76.11              & -                                         & -                                              \\ \cline{4-4}
                                        & FD                                                     & \multicolumn{1}{c|}{}                           &                       & 75.57              & -                                         & -                                              \\
\multirow{-4}{*}{\textbf{FashionMNIST}} & \textbf{FedCache}                                        & \multicolumn{1}{c|}{\multirow{-4}{*}{$A^C_1$, $A^C_2$, $A^C_3$}} & \multirow{-2}{*}{-}   & 77.26              & \textbf{0.08}                             & \textbf{$\times$158.0}                                \\ \hline
                                        & FedDKC                                                 & \multicolumn{1}{c|}{}                           &                       & 44.53              & 4.58                                      & $\times$1.2                                           \\
                                        & FedICT                                                 & \multicolumn{1}{c|}{}                           & \multirow{-2}{*}{$A^S$} & 43.96              & 5.35                                      & $\times$1.0                                           \\ \cline{4-4}
                                        & FD                                                     & \multicolumn{1}{c|}{}                           &                       & 40.40               & -                                         & -                                              \\
\multirow{-4}{*}{\textbf{CIFAR-10}}     & \textbf{FedCache}                                        & \multicolumn{1}{c|}{\multirow{-4}{*}{$A^C_1$, $A^C_2$, $A^C_3$}} & \multirow{-2}{*}{-}   & 41.59              & \textbf{0.05}                             & \textbf{$\times$107.0}                                \\ \hline
                                        & FedDKC                                                 & \multicolumn{1}{c|}{}                           &                       & 44.80               & 4.12                                      & $\times$1.3                                           \\
                                        & FedICT                                                 & \multicolumn{1}{c|}{}                           & \multirow{-2}{*}{$A^S$} & 43.40               & 5.50                                       & $\times$1.0                                           \\ \cline{4-4}
                                        & FD                                                     & \multicolumn{1}{c|}{}                           &                       & 40.76              & -                                         & -                                              \\
\multirow{-4}{*}{\textbf{CINIC-10}}     & \textbf{FedCache}                                        & \multicolumn{1}{c|}{\multirow{-4}{*}{$A^C_1$, $A^C_2$, $A^C_3$}} & \multirow{-2}{*}{-}   & 41.71              & \textbf{0.07}                             & \textbf{$\times$78.6}                                 \\ \hline
\end{tabular}
\label{model-hetero}
\end{table*}

\begin{figure*}[t]
	\centering
	\includegraphics[width=1.0\textwidth]{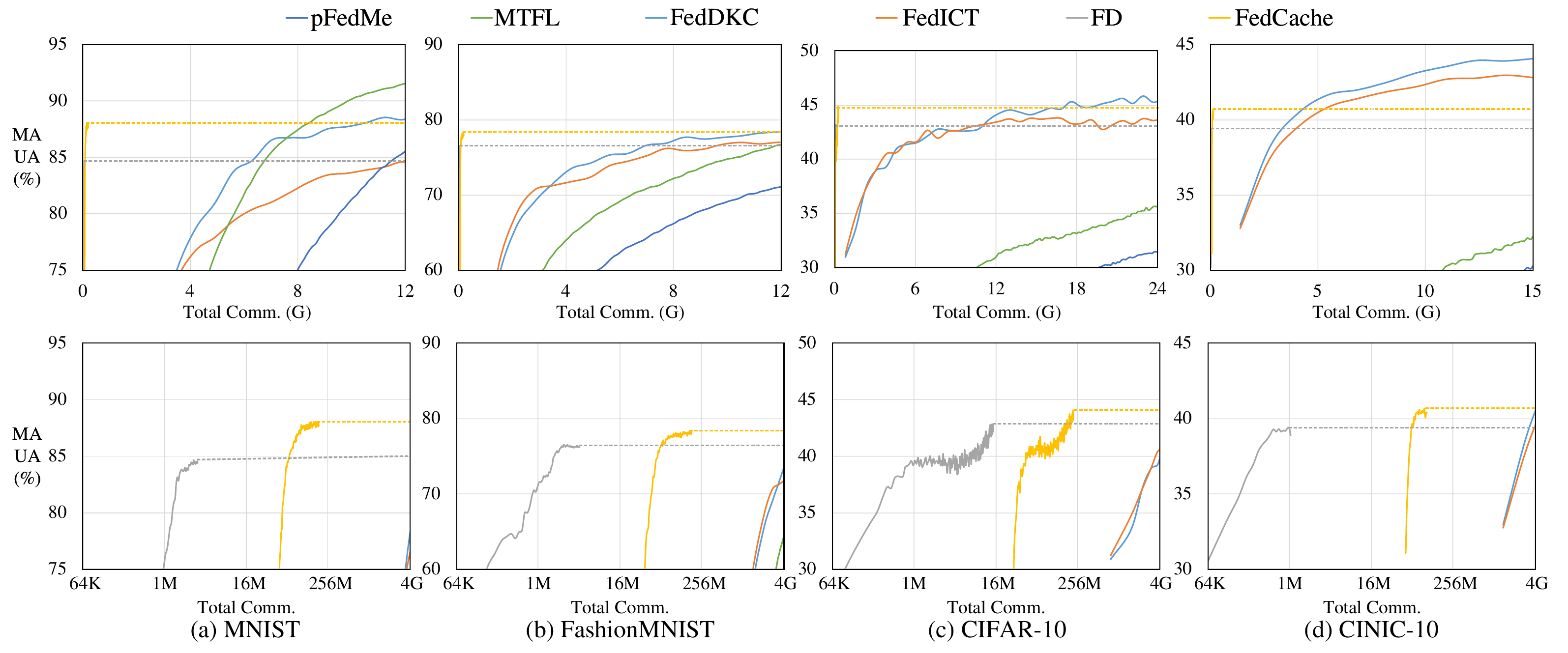}
	\caption{MAUA (\%) per unit of communication overhead in experiments with homogeneous models. Dashed lines indicate the extension of algorithms beyond convergence to the maximum MAUA over communication overheads. The same as below.}
	\label{model-homo-figure}
%\end{figure*}

\vspace {0.8cm}

%\begin{figure*}[t]
	\centering
	\includegraphics[width=1.0\textwidth]{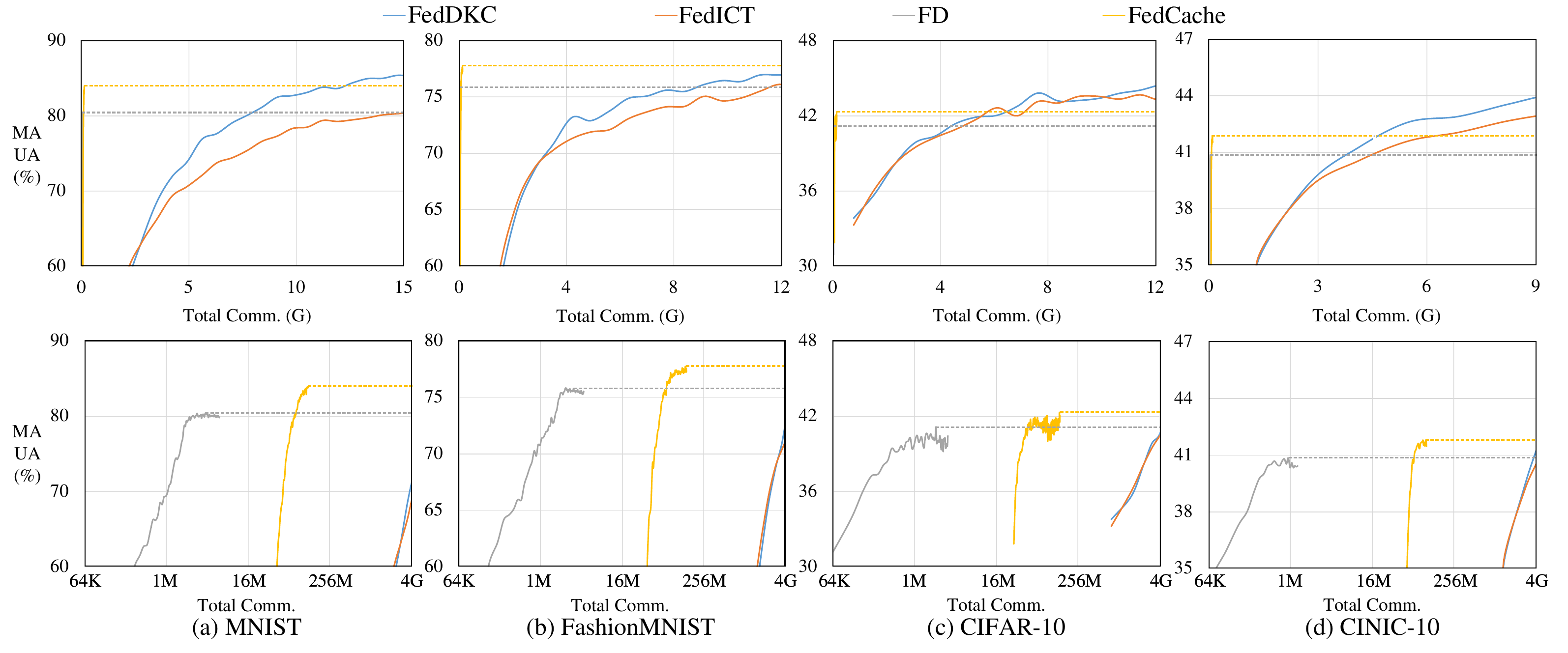}
	\caption{MAUA (\%) per unit of communication overhead in experiments with heterogeneous models.}
	\label{model-hetero-figure}
\end{figure*}

\subsection{Results}
\subsubsection{Performance on Homogeneous Models}
\label{performance-homo}
TABLE \ref{model-homo} displays MAUA and communication overhead of FedCache compared to all considered benchmarks on different datasets, and the MAUA performance per unit of communication overhead is shown in Fig. \ref{model-homo-figure}. As can be seen from TABLE \ref{model-homo}, FedCache achieves 77.71\%, 44.42\%, and 40.45\% MAUA on FashionMNIST, CIFAR-10, and CINIC-10 datasets, respectively, which are is comparable to the considered benchmark algorithms. 
Meanwhile, according to the criteria described in \ref{criterions}, the total communication overhead of FedCache over the three datasets mentioned above are all less than 0.20G, and the speed-up ratios of FedCache are all over $\times78$, which enable the communication efficiency to be much higher than existing methods with previous architectures. This is because FedCache adopts a lightweight communication protocol with only logits and hash values being transferred, and does not transmit model parameters as well as features with relatively large sizes.
Moreover, the efficient communication of FedCache can be further verified in Fig. \ref{model-homo}, where our method exhibits a much steeper convergence curve than FedDKC, FedICT, pFedMe, and MTFL. 
We can also observe in Fig. \ref{model-homo} that compared to communication-efficient FD, FedCache achieves significantly higher MAUA, achieving satisfactory system performance while maintaining communication efficiency orders of magnitude higher than other benchmark algorithms. The reason is that FedCache is an SLIA architecture rather than CLIA, where enriched knowledge can be utilized to obtain significantly more information for on-device model constructive optimization, thus possessing performance superiority.

%\end{table}

%\vspace{8pt}
\subsubsection{Performance on Heterogeneous Models}
TABLE \ref{model-hetero} shows the comparison of FedCache with benchmark algorithms that support model heterogeneity on clients. Likewise, we can conclude that FedCache achieves comparable MAUA to considered benchmarks, but with extremely high communication efficiency due to the elimination of feature transfer compared with FedDKC and FedICT. Similar to section \ref{performance-homo}, FedCache obtains a convergence curve in Fig. \ref{model-hetero-figure} that is capped above FD, which indicates that FedCache gains better system performance than FD. This further confirms the superiority of our proposed FedCache architecture over heterogeneous models.

\section{Ablation Study}
\subsection{Ablation Settings}
%To investigate the effect of different settings on the performance of FedCache, 
In this section, we conduct the ablation study to investigate the impact of three factors on the performance of FedCache: the degree of data heterogeneity, the proportion of local samples, and the number of related samples. All ablation experiments are evaluated on the FashionMNIST dataset, with the same settings adopted for experiments with homogeneous models in section \ref{performance-homo} by default.
The performance of all algorithms is measured by MAUA (\%) in the following subsections.
%Subsequent settings are modified from the aforementioned default settings.

\begin{figure}[!t]
	\centering
	\includegraphics[width=0.5\textwidth]{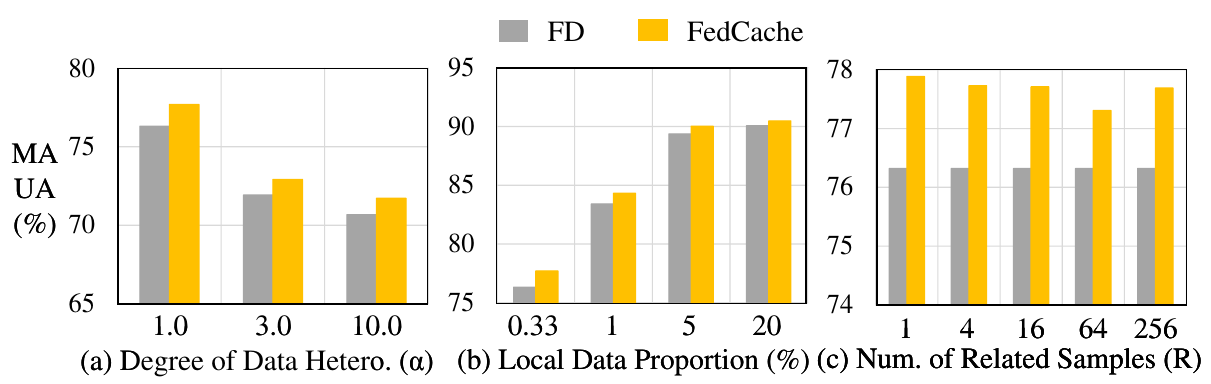}
	\caption{The impact of the degree of data heterogeneity, local data proportion, and the number of related samples on the performance of FedCache.}
	\label{ablation}
\end{figure}

\begin{table}[!t]
	\caption{Performance of FedCache and FD with different $\bm{\alpha}$.}
	\centering
	\setlength{\tabcolsep}{13pt}
	\renewcommand\arraystretch{1.2}
\begin{tabular}{l|ccc}
\hline
\multicolumn{1}{c|}{\multirow{2}{*}{\textbf{Method}}} & \multicolumn{3}{c}{\textbf{Degree of Data Heterogeneity}} \\
\multicolumn{1}{c|}{}                                 & $\bm{\alpha}$\textbf{=1.0}             & $\bm{\alpha}$\textbf{=3.0}             & $\bm{\alpha}$\textbf{=10.0}            \\ \hline
FD                                                    & 76.32             & 71.92             & 70.67             \\
\textbf{FedCache}                                       & \textbf{77.71}    & \textbf{72.92}    & \textbf{71.73}    \\ \hline
\end{tabular}
\label{impact-a}
\end{table}

\begin{table}[t]
	\caption{Performance of FedCache and FD with different average local sample proportions.}
	\centering
	\setlength{\tabcolsep}{10pt}
	\renewcommand\arraystretch{1.2}
\begin{tabular}{l|cccc}
\hline
\multicolumn{1}{c|}{\multirow{2}{*}{\textbf{Method}}} & \multicolumn{4}{c}{\textbf{Average Local Sample Proportion}}      \\
\multicolumn{1}{c|}{}                                 & \textbf{0.33\%}         & \textbf{1\%}            & \textbf{5\%}            & \textbf{20\%}           \\ \hline
FD                                                    & 76.32          & 83.41          & 89.37          & 90.07          \\
\textbf{FedCache}                                       & \textbf{77.71} & \textbf{84.32} & \textbf{90.03} & \textbf{90.47} \\ \hline
\end{tabular}
\label{impact-client-num}
\end{table}

%\vspace {0.8cm}
\subsection{Results}
\subsubsection{Impact of Degree of Data Heterogeneity}
To explore the effect of data heterogeneity on the performance of FedCache, we set the hyper-parameter $\alpha$ to different values $\alpha\in\{1.0, 3.0, 10.0\}$ to control the degree of data heterogeneity, and compare the performance of FedCache with FD, with the results shown in TABLE \ref{impact-a} and Fig. \ref{ablation} (a).
It can be seen that FedCache consistently outperforms FD despite the skewness of local data distributions, reflecting the adaptability of our method to different data environments.

\subsubsection{Impact of Local Data Proportion}
To investigate the performance of FedCache with different percentages of local data to the overall data, we control the number of different local samples to $\{0.33\%, 1\%, 5\%, 20\%\}$ of the whole dataset, and compare the performance of FedCache with FD, with the results shown in TABLE \ref{impact-client-num} and Fig. \ref{ablation} (b). We can observe that the performance of both FD and FedCache improves as the local sample share of clients increases. Still, FedCache always outperforms FD, which confirms the superior performance of our FedCache with varying percentages of local data from a single client.

\subsubsection{Impact of Number of Related Samples}
To evaluate the performance of FedCache with different numbers of related samples, we set $R\in \{1,4,16,64,256\}$ and compare the performance of FedCache with FD with the aforementioned $R$ settings, with the results shown in TABLE \ref{impact-R} and Fig. \ref{ablation} (c). It can be seen that our method consistently outperforms FD in different $R$ settings. This indicates that FedCache is robust to the choice of related samples and can achieve satisfactory performance consistently.

\begin{table}[!t]
    %\begin{table}[!t]
	\caption{Performance of FedCache and FD with different $R$.}
	\centering
	\setlength{\tabcolsep}{10pt}
	\renewcommand\arraystretch{1.2}
\begin{tabular}{l|c}
\hline
\multicolumn{1}{c|}{\textbf{Method}} & \textbf{MAUA(\%)} \\ \hline
FD                                   & 76.32                   \\
\textbf{FedCache ($\bm{R}$=1)}                 & 77.89                   \\
\textbf{FedCache ($\bm{R}$=4)}                 & 77.73                   \\
\textbf{FedCache ($\bm{R}$=16)}                & 77.71                   \\
\textbf{FedCache ($\bm{R}$=64)}                & 77.31                   \\
\textbf{FedCache ($\bm{R}$=256)}               & 77.69                   \\ \hline
\end{tabular}
\label{impact-R}
\end{table}

\begin{table*}[t]
	\caption{Comparison of the computation complexity of PFL architectures. $r$ represents the total communication rounds. $N^P$ represents the number of samples in the public dataset. $F$ represents the scale of transmitted features.}
	\renewcommand\arraystretch{1.5}
	\centering
 \begin{adjustbox}{center}
	\begin{tabular}{c|c|c|c|c|c}
\hline
                                                                       & \textbf{PIA} & \textbf{SLIA-FE} & \textbf{SLIA-PD} & \textbf{CLIA} & \textbf{FedCache} \\ \hline
\textbf{\begin{tabular}[c]{@{}c@{}}End\\ Devices\end{tabular}} & $r{N^k} \cdot {\rm{O}}({W^k})$               & $r{N^k} \cdot {\rm{O}}({W^k})$               & $r({N^k} + {N^P}) \cdot {\rm{O}}({W^k})$               & $r{N^k} \cdot {\rm{O}}({W^k})$               & $r{N^k} \cdot {\rm{O}}({W^k})$              \\ \hline
\textbf{\begin{tabular}[c]{@{}c@{}}Edge\\ Server\end{tabular}}        & $rK \cdot {\rm{O}}({W^S})$               & $r\sum\limits_{k = 1}^K {{N^k}}  \cdot {\rm{O}}({W^S})$               & $r{N^P}K\cdot {\rm{O}}(C)$               & $rK\cdot {\rm{O}}({C^2})$               & 
$\begin{array}{l}
R\sum\limits_{k = 1}^K {{N^k}} \log \sum\limits_{k = 1}^K {{N^k}}  \cdot {\rm{O}}(F)\\
 + rR\sum\limits_{k = 1}^K {{N^k}}  \cdot {\rm{O}}(C)
\end{array}$
\\ \hline
\end{tabular}
\end{adjustbox}
	\label{cmp-complex}
\end{table*}

\section{Discussion}
\subsection{Analysis on Computation Complexity}
We compare the communication complexity of FedCache with other PFL architectures in TABLE \ref{cmp-complex}.
On the device side, FedCache has the identical computation complexity as PIA, SLIA-FE and CLIA, since their computation all mainly focuses on on-device models' forward propagation on local data.
As there is no need for local training based on public datasets, FedCache has lower computation complexity on the device side compared to SLIA-PD.
On the server side, the computation overhead of FedCache mainly consists of establishing relations among samples through $R$-nearest neighbors retrieval and integrating relevant knowledge of a given sample index in each round.
The server-side computation complexity comparison between FedCache and PIA depends on the scale of model parameters and the average number of local samples per client. 
In our experiments on four datasets with $R$ = 16, $r>100$, hundreds of samples held on a single client and the parameter size $>50K$, the server-side computation complexity of FedCache is much smaller than that of PIA.
Since FedCache converges better than PIA in empirical experiments, the superiority of computation overhead  of FedCache over PIA will be pronounced in reality.
In addition, we claim that the server-side computation complexity of FedCache is much smaller than that of SLIA-FE and SLIA-PD.
The reason for the former is that FedCache doesn't require forward propagation on server-side model training for each sample. 
While the reason for the latter is that the average size of private data per client is much smaller than that of public datasets in practice.
Although CLIA achieves relatively-low server-side computation complexity over FedCache, it pays the price of significantly reduced knowledge enrichment, resulting in poor performance confirmed by empirical experiments in section \ref{exper}, so FedCache still possesses an irreplaceable superiority over CLIA in terms of balancing computation overhead and system performance.

\subsection{Limitations}
We analyze that the limitations of FedCache are threefold. 
One limitation of FedCache is that it conducts knowledge distillation on device-side models only based on knowledge associated with local samples, but neglects knowledge learning for global generalization.
As a result, it is only suitable for personalization tasks rather than general tasks that require global generalization capabilities.
The generalization performance of FedCache can be improved when introducing additional information, such as partial global parameters or global public data.
Another limitation is that we apply our method only to conventional image classification problems in our experiments, and additional research on data encoding strategies, hash correlation measures for serialized data and other non-image structured data are also meaningful for FedCache.
\textcolor{black}{In addition, FedCache cannot support PFL with dynamism and continuity data, while end devices may continuously generate new data that requires real-time processing and analysis \cite{savaglio2021simulation}.}
By considering and addressing the above limitations, FedCache can further enhance its effectiveness in a wider range of applications.

\section{Conclusion}
In this paper, we propose FedCache, a novel federated learning architecture tailored for personalized edge intelligence. FedCache designs a knowledge cache on the server for storing newly-extracted knowledge uploaded by clients and fetching correlatively personalized knowledge from samples with similar hashes to the specified private data. 
On this basis, ensemble distillation is performed on device-side local models for personalized constructive optimization.
To our best knowledge, FedCache is the first architecture for personalized federated learning that enables sample-grained logits interaction without features transmission or public datasets. Empirical experiments show that FedCache achieves comparable accuracy with state-of-the-art personalized federated learning methods with various architectures, meanwhile reducing communication costs by two orders of magnitude.
%To our best knowledge, FedCache is the first sample-grained logits interaction-based architecture without features transmission and public datasets for personalized federated learning. Empirical experiments demonstrate that FedCache achieves comparable performance with state-of-the-art personalized federated learning approaches in terms of accuracy, and meanwhile increasing communication efficiency by two orders of magnitudes.

\section*{Acknowledgments}
We thank Prof. Anfu Zhou from Beijing University of Posts and Telecommunications, Tian Wen, Quyang Pan, Xujing Li, Chungang Lin from the Institute of Computing Technology, Chinese Academy of Sciences, and Yuhan Tang, Aoxu Zhang from Beijing Jiaotong University for inspiring suggestions.
\ifCLASSOPTIONcaptionsoff
  \newpage
\fi

\bibliographystyle{IEEEtran}

\vspace{-33pt}
\begin{IEEEbiography}[{\includegraphics[width=1in,height=1.25in,clip,keepaspectratio]{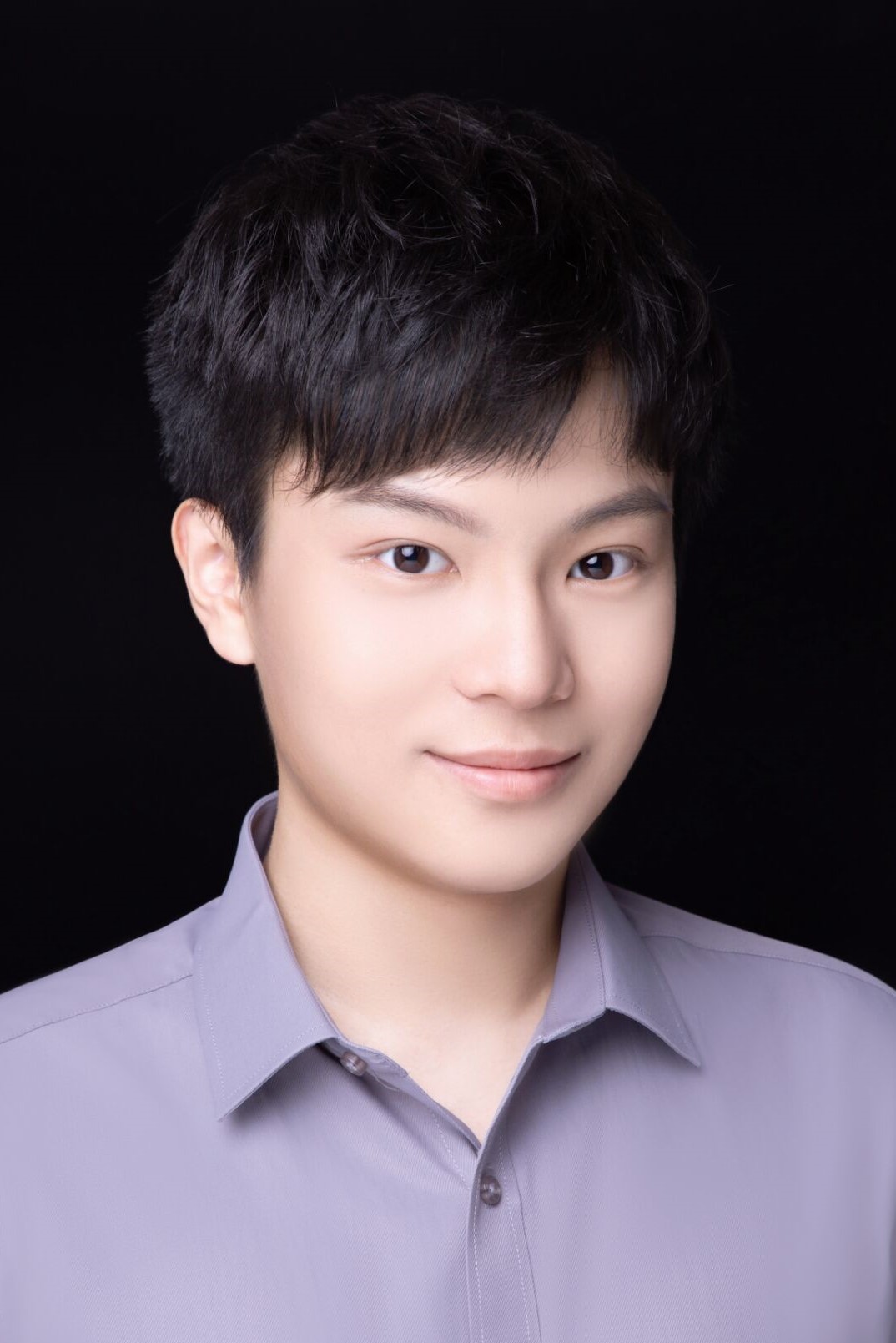}}]{Zhiyuan Wu}
	(Member, IEEE) is currently a research assistant with the Institute of Computing Technology, Chinese Academy of Sciences (ICT, CAS). He has contributed several technical papers to top-tier conferences and journals as the first author in the fields of computer architecture, computer networks, and intelligent systems, including IEEE Transactions on Parallel and Distributed Systems (TPDS), IEEE Transactions on Mobile Computing (TMC), IEEE International Conference on Computer Communications (INFOCOM), and ACM Transactions on Intelligent Systems and Technology (TIST). He has served as a technical program committee member or a reviewer for over 10 conferences and journals, and was invited to serve as a session chair for the International Conference on Computer Technology and Information Science (CTIS).
    He is a member of IEEE, ACM, the China Computer Federation (CCF), and is granted the President Special Prize of ICT, CAS. His research interests include federated learning, mobile edge computing, and distributed systems.

 %His commitment to the academic community is further exemplified by his active role as a technical program committee member and reviewer for over ten conferences and journals. Additionally, he is an esteemed member of the Distributed Computing and Systems Committee at the China Computer Federation (CCF). His research is primarily focused on federated learning, mobile edge computing, and distributed systems, areas in which he continues to explore and innovate.
 
 %is currently a research assistant with the Institute of Computing Technology, Chinese Academy of Sciences. He has published several technical papers as the first author in top-tier conferences and journals related to computer architecture, computer networks, and intelligent systems, including TPDS, TMC, INFOCOM, TIST, etc. He has served as a technical program committee member or reviewer for 10+ conferences and journals, and is also a member of the Distributed Computing and Systems Committee, China Computer Federation (CCF). His research interests include federated learning, mobile edge computing, and distributed system.
\end{IEEEbiography}

\begin{IEEEbiography}[{\includegraphics[width=1in,height=1.25in,clip,keepaspectratio]{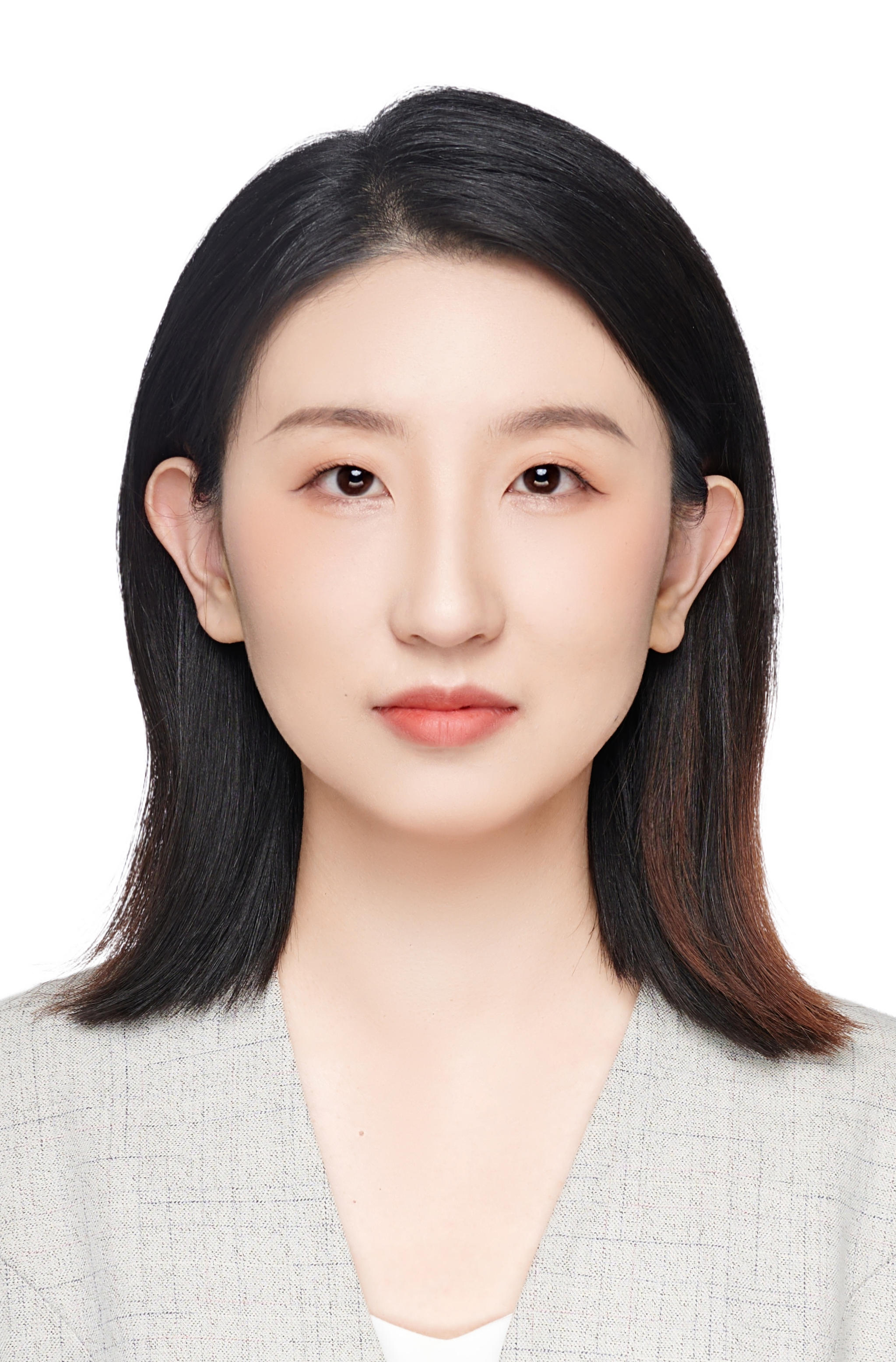}}]{Sheng Sun} is currently an associate professor at the Institute of Computing Technology, Chinese Academy of Sciences. She received her bachelor's degree from Beihang University, and her Ph.D. from the Institute of Computing Technology, Chinese Academy of Sciences. Dr. Sun has led or executed 5 major funded research projects and published over 20 technical papers in journals and conferences related to computer network and distributed systems, including IEEE Transactions on Parallel and Distributed Systems (TPDS), IEEE Transactions on Mobile Computing (TMC), and IEEE International Conference on Computer Communications (INFOCOM). Her research interests include federated learning, edge intelligence, and privacy computing.
	
 %received her B.S. and Ph.D degrees in computer science from Beihang University, China, and the University of Chinese Academy of Sciences, China, respectively. She is currently an associate professor at the Institute of Computing Technology, Chinese Academy of Sciences, Beijing, China. Her current research interests include federated learning, mobile computing and edge intelligence. 
\end{IEEEbiography}

\begin{IEEEbiography}[{\includegraphics[width=1in,height=1.25in,clip,keepaspectratio]{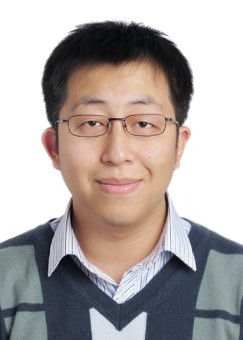}}]{Yuwei Wang}
	(Member, IEEE) received his Ph.D. degree in computer science from the University of Chinese Academy of Sciences, Beijing, China. He is currently an associate professor at the Institute of Computing Technology, Chinese Academy of Sciences. He has been responsible for setting over 30 international and national standards, and also holds various positions in both international and national industrial standards development organizations (SDOs) as well as local research institutions, including the associate rapporteur at the ITU-T SG16 Q5, and the deputy director of China Communications Standards Association (CCSA) TC1 WG1. His current research interests include federated learning, mobile edge computing, and next-generation network architecture.
\end{IEEEbiography}

\begin{IEEEbiography}[{\includegraphics[width=1in,height=1.25in,clip,keepaspectratio]{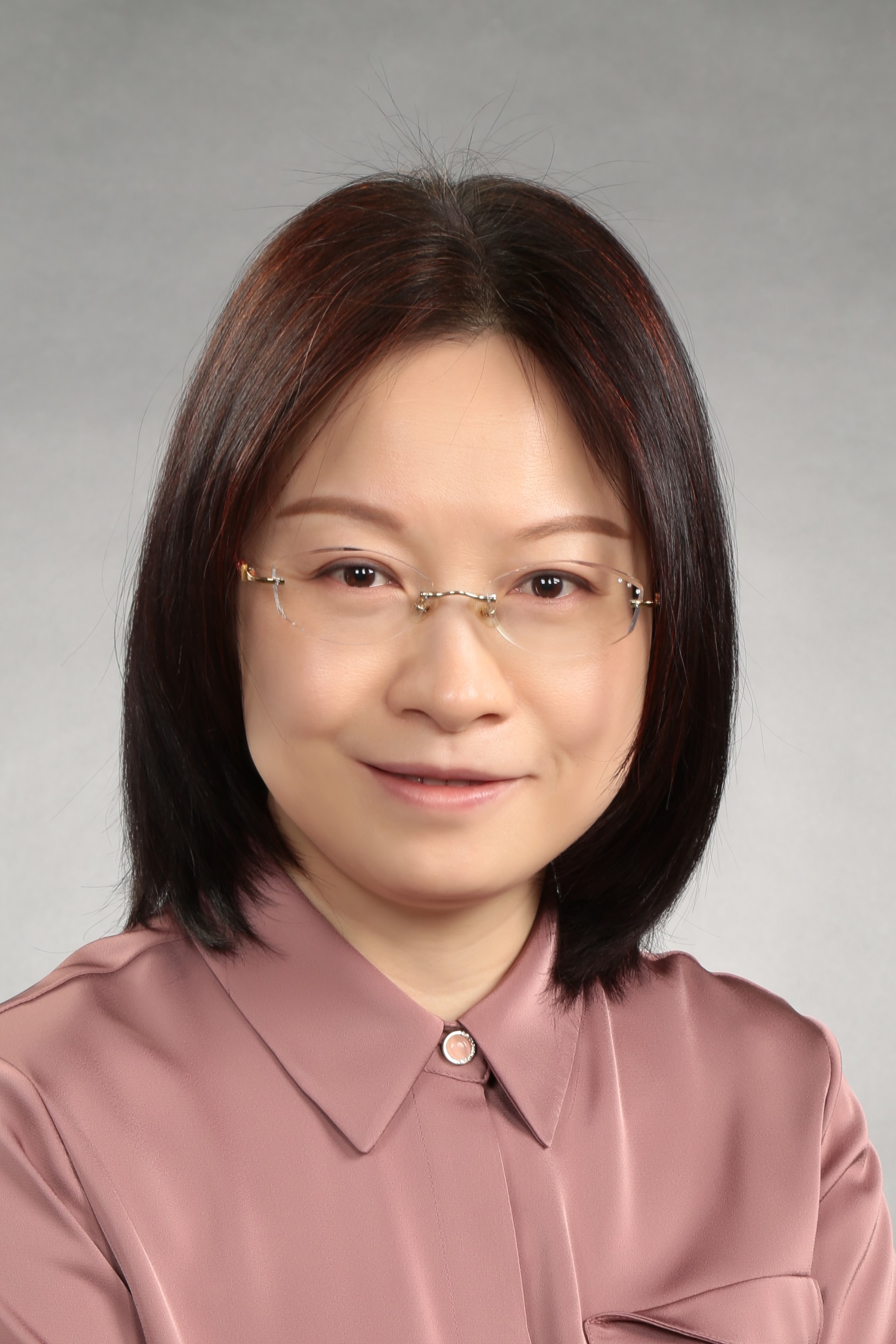}}]{Min Liu}
	(Senior Member, IEEE) received her Ph.D degree in computer science from the Graduate University of the Chinese Academy of Sciences, China. Before that, she received her B.S. and M.S. degrees in computer science from Xi’an Jiaotong University, China. She is currently a professor at the Institute of Computing Technology, Chinese Academy of Sciences, and also holds a position at the Zhongguancun Laboratory. Her current research interests include mobile computing and edge intelligence.
\end{IEEEbiography}

\begin{IEEEbiography}[{\includegraphics[width=1in,height=1.25in,clip,keepaspectratio]{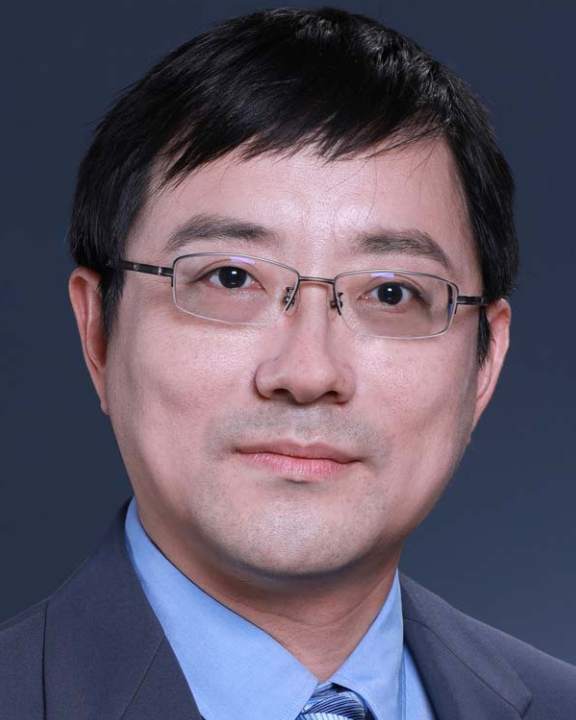}}]{Ke Xu}
	(Senior Member, IEEE) received the Ph.D. degree from the Department of Computer Science and Technology, Tsinghua University, Beijing, China. He serves as a Full Professor at Tsinghua University. He has published more than 200 technical papers and holds 11 U.S. patents in the research areas of next-generation internet, blockchain systems, the Internet of Things, and network security. He is a member of ACM. He served as the Steering Committee Chair for IEEE/ACM IWQoS.
He has guest-edited several special issues in IEEE and Springer journals. He is the Editor of IEEE Internet of Things Journal.
\end{IEEEbiography}

\begin{IEEEbiography}[{\includegraphics[width=1in,height=1.25in,clip,keepaspectratio]{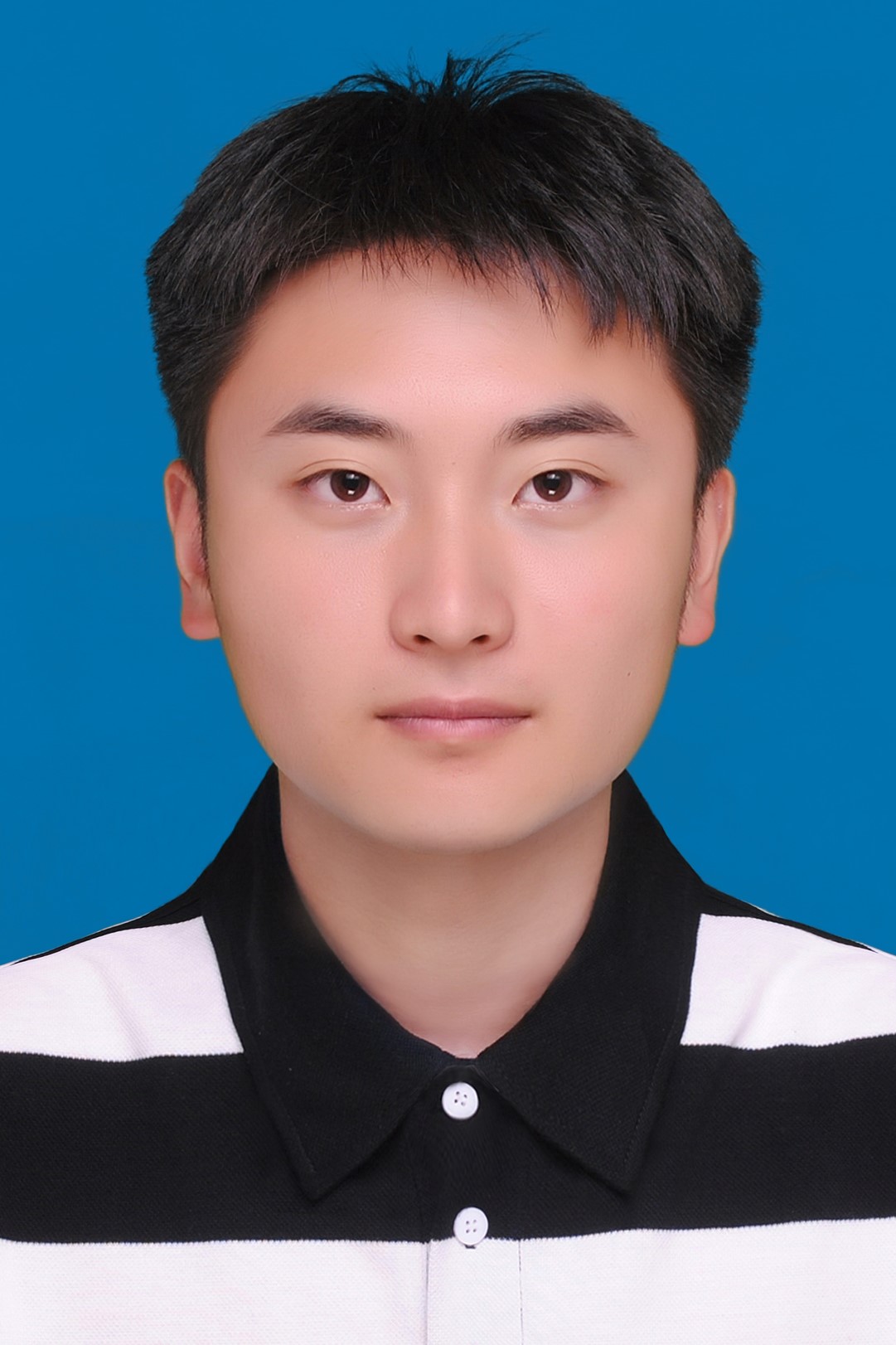}}]{Wen Wang} 
	is currently a master’s candidate with the Institute of Computing Technology, Chinese Academy of Sciences. Before that, he received his bachelor degree with honor at Xidian University. His research interests include network security, in-network computing and information retrieval.
\end{IEEEbiography}

\begin{IEEEbiography}[{\includegraphics[width=1in,height=1.25in,clip,keepaspectratio]{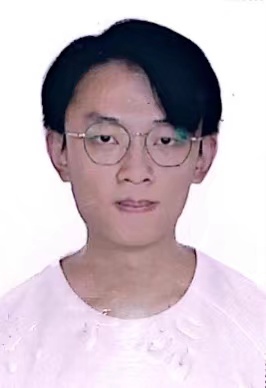}}]{Xuefeng Jiang}
	is currently a Ph.D candidate with the Institute of Computing Technology, Chinese Academy of Sciences. Before that, he received his bachelor degree with honor at Beijing University of Posts and Telecommunications. His research interests include distributed optimization and machine learning.
\end{IEEEbiography}

\begin{IEEEbiography}[{\includegraphics[width=1in,height=1.25in,clip,keepaspectratio]{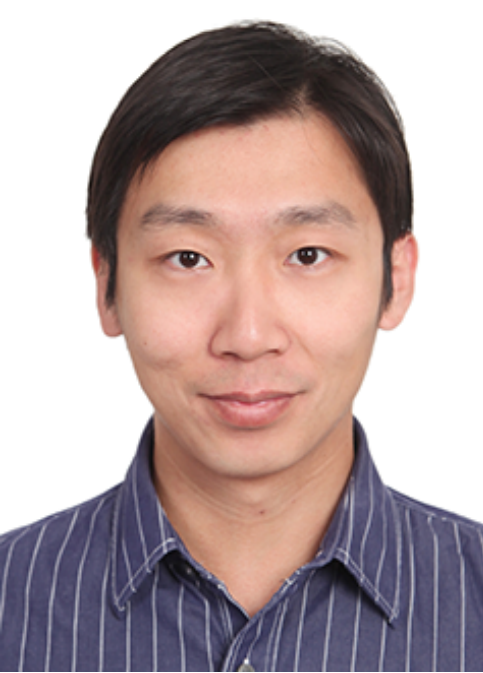}}]{Bo Gao} (Member, IEEE)
	received his M.S. degree in electrical engineering from the School of Electronic Information and Electrical Engineering at Shanghai Jiaotong University, Shanghai, China in 2009, and his Ph.D. degree in computer engineering from the Bradley Department of Electrical and Computer Engineering at Virginia Tech, Blacksburg, USA in 2014. He was an Assistant Professor with the Institute of Computing Technology at Chinese Academy of Sciences, Beijing, China from 2014 to 2017. He was a Visiting Researcher with the School of Computing and Communications at Lancaster University, Lancaster, UK from 2018 to 2019. He is currently an Associate Professor with the School of Computer and Information Technology at Beijing Jiaotong University, Beijing, China. He has directed a number of research projects sponsored by the National Natural Science Foundation of China (NSFC) or other funding agencies. He is a member of IEEE, ACM, and China Computer Federation (CCF). His research interests include wireless networking, mobile/edge computing, multiagent systems, and machine learning.
\end{IEEEbiography}

% \begin{IEEEbiography}[{\includegraphics[width=1in,height=1.25in,clip,keepaspectratio]{htl.jpg}}]{Tianliu He}
% 	is currently a master’s candidate with the Institute of Computing Technology, Chinese Academy of Sciences. Before that, he received his bachelor's degree with honors at Huazhong University of Science and Technology. His research interests include federated learning and machine learning.
% \end{IEEEbiography}

\begin{IEEEbiography}[{\includegraphics[width=1in,height=1.25in,clip,keepaspectratio]{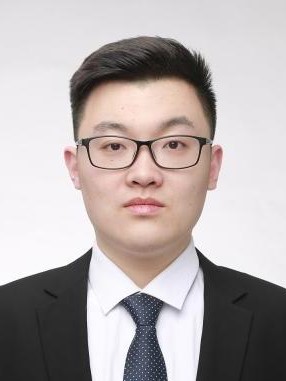}}]{Jinda Lu}
	is currently a master’s candidate with the School of Information Science and Technology, University of Science and Technology of China. Before that, he received his bachelor's degree with honor at Jilin University. His research interests include artificial intelligence, pattern recognition, and knowledge distillation.
\end{IEEEbiography}

\vfill
\newpage
\appendices
\onecolumn
\end{document}